\documentclass[a4paper,fleqn,usenatbib]{mnras}

\usepackage{newtxtext,newtxmath}

\usepackage[T1]{fontenc}
\usepackage{ae,aecompl}

\usepackage{graphicx}	
\usepackage{amsmath}	
\usepackage{amssymb}	
\usepackage{verbatim}

\usepackage{natbib, hyperref}
\usepackage{amssymb, amsmath}
\usepackage{color}
\usepackage{times}
\usepackage{rotating}
\usepackage{lscape}
\usepackage{enumerate}

\long\def\symbolfootnote[#1]#2{\begingroup%
\def\thefootnote{\fnsymbol{footnote}}\footnote[#1]{#2}\endgroup}

\newcommand{\tr}[1]{\textcolor{black}{#1}}
\newcommand{\trr}[1]{\textcolor{red}{#1}}

\title[Proper-Motion Cleaned HST Catalogs of BSSs in GCs]{Blue Straggler Star Populations in Globular Clusters: II. Proper-Motion Cleaned HST Catalogs of BSSs in \tr{38} Galactic GCs\thanks{Based on observations made with the NASA/ESA Hubble Space Telescope, obtained from the Data Archive at the Space Telescope Science Institute, which is operated by the Association of Universities for Research in Astronomy, Inc., under NASA contract NAS 5-26555. These observations are associated with program GO-10775 and GO-13297.}}

\author[Simunovic \& Puzia]{
Mirko Simunovic$^{1,\thanks{E-mail: msimunov@astro.puc.cl}}$
\& Thomas H. Puzia$^{1,\thanks{E-mail: tpuzia@astro.puc.cl}}$
\\
$^{1}$Institute of Astrophysics, Pontificia Universidad Cat\'olica de Chile, Av. Vicu\~na Mackenna 4860, 7820436 Macul, Santiago, Chile
}

\date{Accepted XXX. Received YYY; in original form \today}

\pubyear{2016}

\begin{document}
\label{firstpage}
\pagerange{\pageref{firstpage}--\pageref{lastpage}}
\maketitle

\begin{abstract}
We present new Blue Straggler Star (BSS) catalogs in \tr{38} Milky Way globular clusters (GCs) based on multi-passband and multi-epoch treasury survey data from the Hubble Space Telescope. We measure precise astrometry and relative proper motions of stars in all target clusters and performed a subsequent cluster membership selection. We study the accuracy of our proper motion measurements using estimates of central velocity dispersions and find very good agreement with previous studies in the literature. Finally, we present a homogeneous BSS selection method, that expands the classic BSS selection parameter space to more evolved BSS evolutionary stages. We apply this method to the proper-motion cleaned GC star catalogs in order to define proper-motion cleaned BSS catalogs in all \tr{38} GCs, which we make publicly available to enable further study and follow-up observations.
\end{abstract}

\begin{keywords}
stars: blue stragglers -- globular clusters: general -- proper motions
\end{keywords}


\section{Introduction}
The current understanding of the evolution of globular clusters (GCs) is in part determined by the dynamical interactions between the member stars and the evolution of the large-scale gravitational potential.~It is clear that the denser regions in GCs are not collisionless systems and therefore two-body relaxation processes are expected to affect global properties, such as the rate of close stellar interactions inducing binary hardening leading to mass-transfer and/or merger events.~Blue Straggler Stars (BSSs) are likely formed in such interactions \citep{mccrea64, hillsday76}, and it has been shown that the present BSS population and its abundance ratio scales with GC structural parameters \tr{\citep{leigh07, piot04, knig09}} giving constraints to the likely formation scenario \tr{\citep{leigh11, leigh13, davies15}}.\,At the same time, the dynamical evolution of GCs is driving much of the initial conditions, and BSSs are in fact good tracers of their current dynamical state \citep{ferr12, ferr15}.\,In this broader context it is \tr{necessary} that we study BSS populations in the most general approach, by looking at them and their properties in multiple galactic GCs.

This approach led us to start looking at the dynamical properties of BSSs in multiple GCs. In particular, \tr{in a previous paper} we studied the radial velocities and rotational velocities of BSSs in NGC\,3201, NGC\,6218 and NGC\,5139 ($\omega$Centauri) and found interesting results regarding their dynamical properties \citep{sipu14}. That study showed an apparent central segregation of fast rotating BSSs, which were preferentially located within one core radius of their parent cluster. This result was confirmed independently by \cite{mucc14} for the case of $\omega$Cen, who also found a peak in the fast rotating BSS fraction at the inner regions of $\omega$Cen. This type of observation suggests that fast rotating BSSs form preferentially in the inner regions of GCs, where the higher rates of dynamical interactions may facilitate formation conditions that favor high angular momentum transfer.~\tr{Such fast rotating BSSs observed in the inner cluster regions may be considered proxies for young BSSs, if these spin-down over time due to strong magnetic braking following their formation.}~Another clue that points towards this scenario is the finding of collision product BSSs in the inner regions of GCs \citep{ferr09,simu14}, that seem to be coeval and produced by such environments of higher rates of dynamical interactions.~Nevertheless, although the crucial interplay between BSS formation and GC dynamical evolution has recently gained enormous attention, a full dynamical characterization of BSSs is still lacking in the literature.~In this work, we present a large effort to obtain proper motion cleaned catalogs of BSSs in several Galactic GCs. 

The paper is organized as follows. Section~\ref{txt:data} presents the data and data reduction, while section~\ref{txt:propm} describes the methods adopted for proper motion measurements and cluster member selection. Section~\ref{txt:anls} presents the analysis of the proper motion accuracy and the selection of BSS candidates.~We summarize our work in Section~\ref{txt:sum}.

\section{Observations and Data Reduction}
\label{txt:data}
This work is based on Hubble Space Telescope (HST) observations of Galactic GCs which come from two large photometric Surveys. Firstly, "{\it The ACS Globular Cluster Survey}" (PI: Ata Sarajedini, HST Program 10775) which provides us with fully-calibrated photometric catalogs for the inner regions of GCs in the F606W and F814 filters, available on the ACS/WFC camera. Secondly, "{\it The HST Legacy Survey of Galactic Globular Clusters: Shedding UV Light on Their Populations and Formation}" (PI: Giampaolo Piotto, HST Program 13297), from which we obtain imaging data for the inner regions of GCs in the F275W, F336W, and F436W filters available on the WFC3/UVIS camera.

\subsection{ACS/WFC Photometry}
The ACS photometric catalogs come from the HST/ACS Galactic Globular Cluster Survey \citep{sara07}.~It consists of $\sim$30 min.~exposures in the F606W ($\sim\!V$) and F814W ($\sim\!I$) bands for the central $\sim\!3.4\arcmin\times3.4\arcmin$ fields of 74 GCs.~The photometry available in the online catalogs has been corrected to account for updated HST/ACS WFC zero points and calibrated in the Vega photometric system.~These catalogs provide high quality photometry down to $\sim\!6$\,mag below the main-sequence turn-off of most GCs. \tr{For quality purposes, we filter out all detections that have a QFIT value larger than 0.2 in the F606W or F814W photometry. See \cite{sara07} for further details on this value. }

\subsection{WFC3/UVIS Photometry}\label{sec:wfc3phot}
For the purpose of this work, we limit ourselves to the F336W filter images alone, given that this filter has usually the most amount of frame exposures per cluster field, which serves our main goal of obtaining accurate astrometric measurements. The original data is comprised of FLT images downloaded from the Mikulski Archive for Space Telescopes (MAST)\footnote{https://archive.stsci.edu}. These FLT images are the calibrated and flat-field corrected by the automatic calibration pipeline CALWF3. The procedure that we apply for the construction of the final photometric catalogs is explained in the following subsections.

\subsubsection{Charge-Transfer Efficiency Correction}\label{sec:ctecorrection}
It is well known that, when in orbit, the HST detectors suffer cumulative radiation damage. As a consequence, this produces charge traps that affect the movement of electrons during detector read-out. The observed effect of this diminished charge-transfer efficiency (CTE) is that point sources leave traces of charge in the direction of the amplifiers, affecting more pixels the further away from the amplifiers. Ideally, one would want all the trace counts that "leaked" from a point source to be "put back" to their original position on the detector. For such a correction, we apply the \texttt{wfc3uv\_ctereverse} script, available from the STScI website, that converts the FLT files into FLC files.~This script reverses the CTE effect, but does not offer a perfect correction, as the measured centroids of the stellar sources suffer small systematic offsets that are dependent on the source location across the chip. However, if understood this does not affect our results given that, as we shall see later, this uncertainty in the centroid is much smaller than the proper motion dispersion threshold we typically use for the cluster membership selection function.

\subsubsection{Source Detection and Flux Measurements}
The corresponding point-source photometry is performed on the FLC images using the standard \texttt{img2xym\_wfc3uv} script provided by STScI. The software typically outputs $\sim\!200$k detections per frame. We adopt a higher-limit of 0.3 for the {\sc QFIT} parameter, which records the fractional disagreement between the model and the image pixels, in order to filter out most of the noisy detections. \tr{Compared to the ACS photometry, this constraint has to be more relaxed given the shallower photometry of the WFC3 dataset.} \\

\subsubsection{Geometric-distortion Corrections}
The WFC3 UVIS detector is tilted at $\sim\!21^{\circ}$ about one of its diagonals, producing a projected rhomboidal elongation of $\sim\!7\%$. This in turn changes the plate scale across the field. More precisely, the sky covered by a UVIS pixel varies by about $\sim$7\% from corner to corner over the full field of view. Hence, the raw source coordinates obtained from the photometry are calibrated with the \texttt{WFC3\_UVIS\_gc} script in order to remove the corresponding geometric distortion effects.~This is our final calibration and, from here on, all WFC3/UVIS catalogs mentioned in the paper are implicitly the result of this procedure.

\begin{figure}[ht!]
\centering
 \includegraphics[width=\linewidth, bb=60 60 890 890]{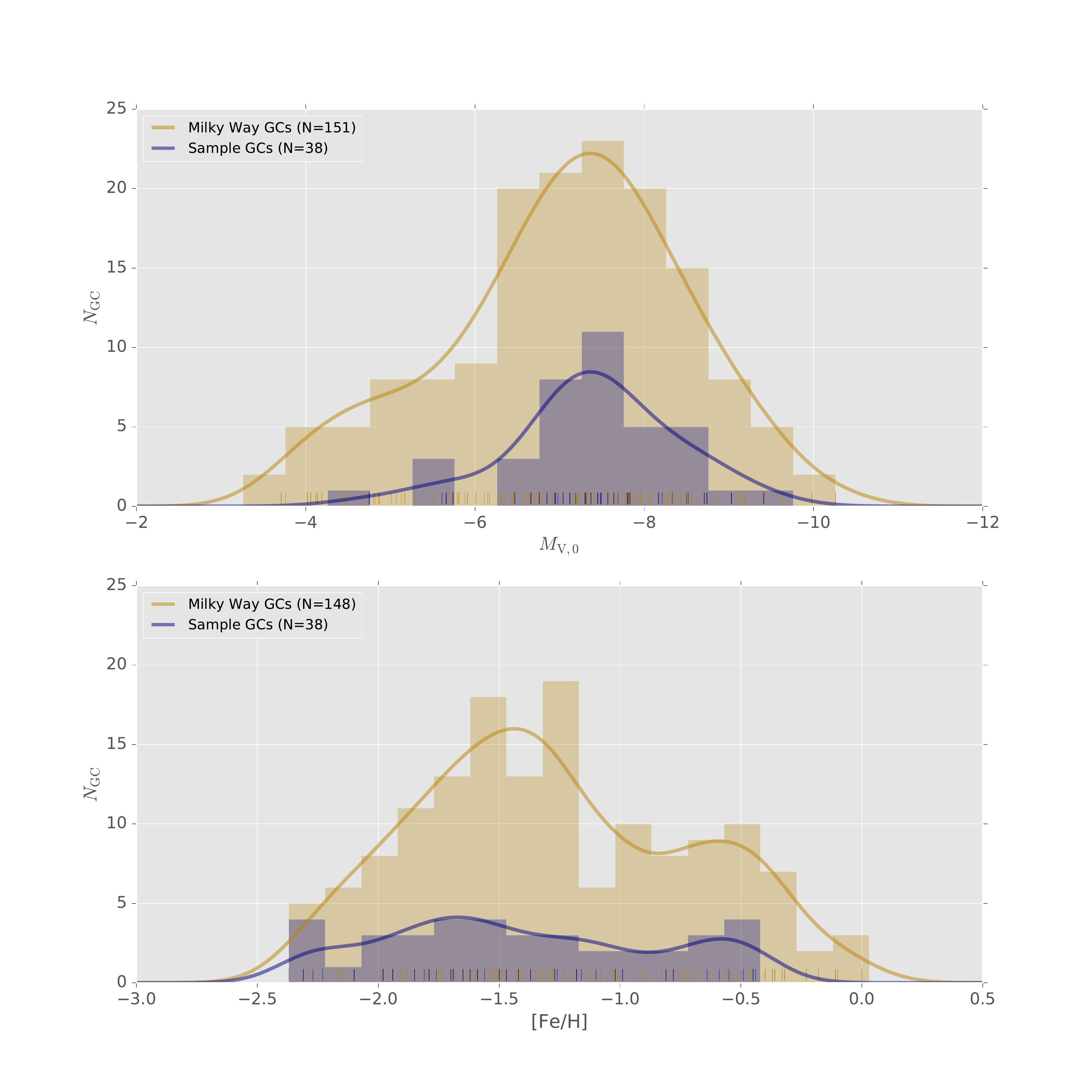} 
 \caption{\tr{Luminosity and metallicity distribution functions of our 38 sample GCs in comparison with the total Milky Way GC system. The solid curves show non-parametric probability density estimates for each distribution. All values were taken from \protect\cite{har10}.} }
 \label{fig:distrib}
\end{figure}

 \tr{The final selection of the sample is mostly determined by the quality of the WFC3/UVIS imaging data. In particular, we decide to remove from the sample a small number of catalogs because of saturation affecting the position measurements of bright stars. The final target sample consists of 38 GCs, and their main properties are listed in Table~\ref{tab:sample}, while Figure~\ref{fig:distrib} shows the luminosity and metallicity distribution of the sample in relation to the entire Milky Way GC system.}    

\section{Measuring Proper Motions}
\label{txt:propm}
One of the advantages of combining the optical and near-UV catalogs is the possibility of measuring proper motions.~Indeed, the two \tr{datasets} used in our work were taken approximately 7 years apart, therefore providing a long enough period for the proper motions to be detectable. The series of steps taken to obtain proper motions can be summarized as follows: 
\begin{enumerate}[{(1)}]
\item For a given WFC3/UVIS filter, we find linear coordinate transformations for all different exposures (between 2 and 6 depending on the cluster) of the same cluster field and set them all to the coordinate frame of an arbitrary exposure.
 \item We do a cross-match and keep only the sources detected in all exposures. We then calculate the average coordinate for each source and repeat (1) using the average coordinates as master the frame. 
 \item We calculate again the average coordinate for each source and and keep them as final coordinates of that cluster field. 
 \item We find linear transformations from the WFC3 master coordinates to the ACS reference frame. 
 \item We cross-match both catalogs and, based on preliminary proper motions, select high-likelihood members and use them to find a new transformation solution. 
 \item We use this final transformation to map the WFC3 catalog coordinates into the ACS master reference frame. 
 \item We cross-match again both catalogs and measure the proper motions in pixel units directly as the difference between the two coordinate sets.
\end{enumerate}
The details of this procedure are fully described in the following subsections.

\begin{figure}
\centering
 \includegraphics[width=\linewidth,bb=30 40 530 410]{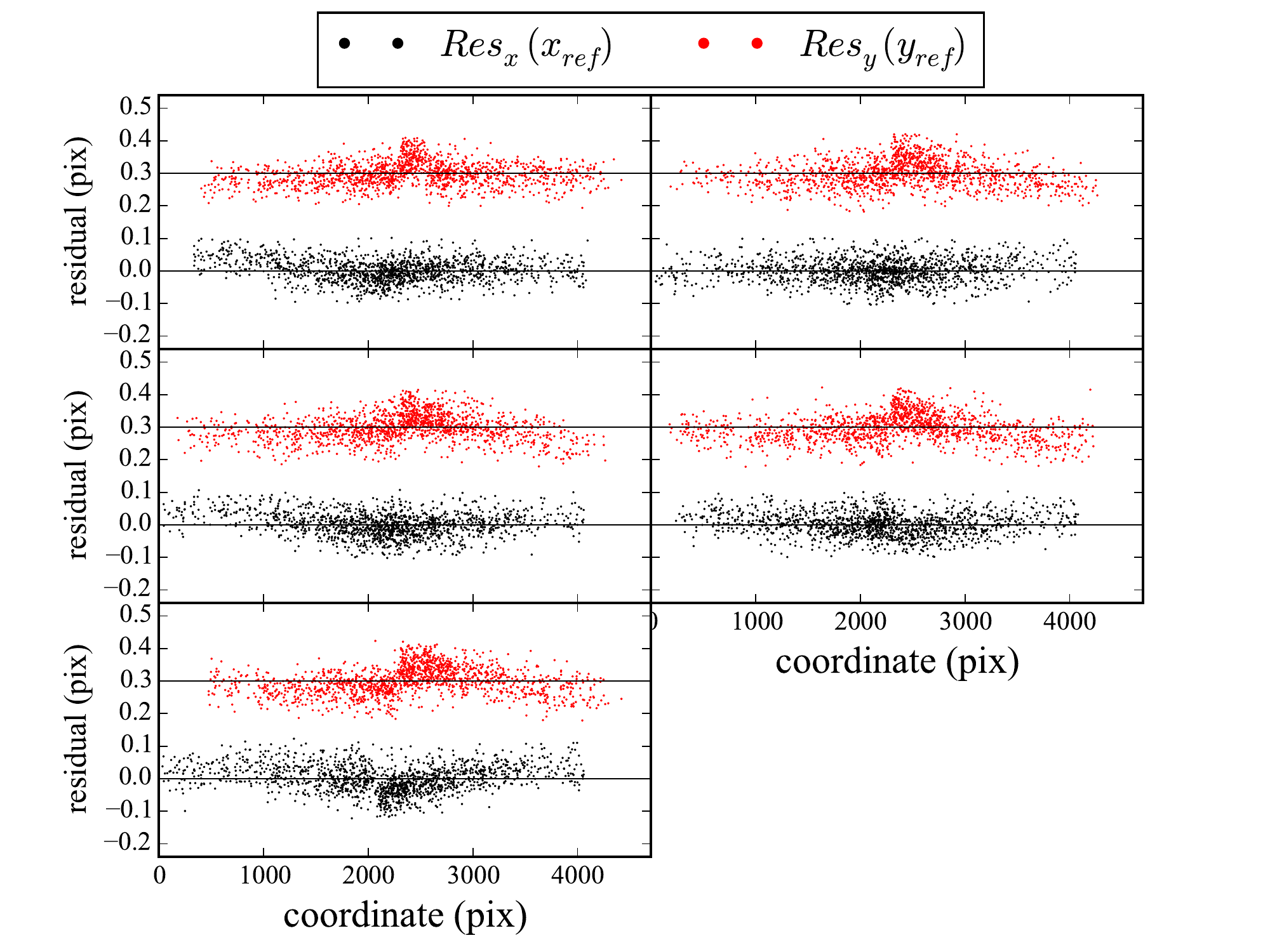} 
 \caption{Coordinate transformation residuals obtained with \texttt{ccmap}. We show in each panel the \tr{corresponding} residuals in the mapping of each individual NGC\,6717 WFC3 exposure into the reference frame exposure. Black and red points show the residual in $X$ as a function of reference $X$ coordinate and the residual in $Y$ as a function of reference $Y$ coordinate, respectively. The residuals in $Y$ have been shifted for clarity. \label{ccmapsol1}}
\end{figure}

\begin{figure}
\centering
 \includegraphics[width=\linewidth,bb=30 40 530 410]{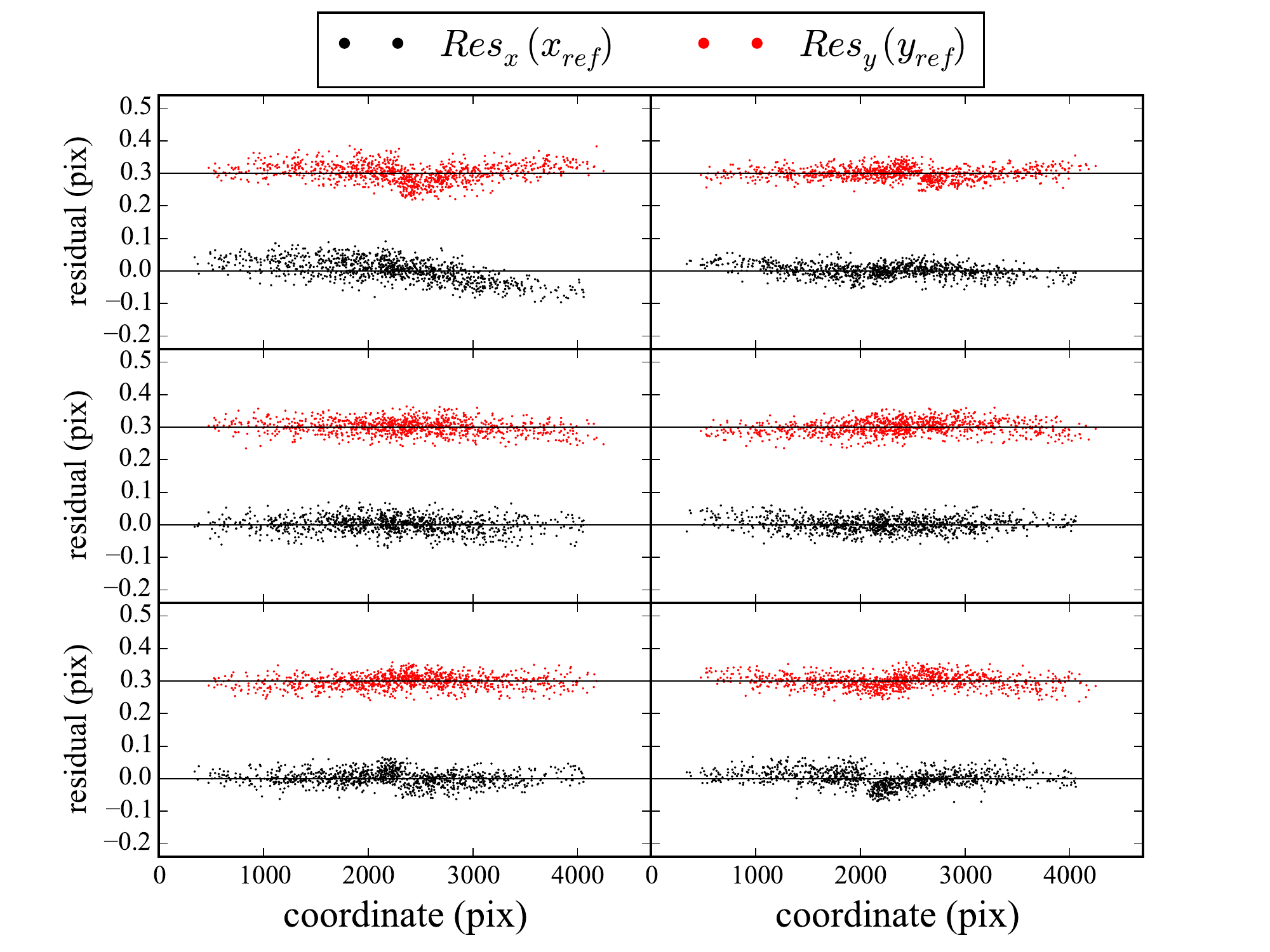} 
 \caption{Same as Fig.~\ref{ccmapsol1} but using the average $XY$ coordinates from all exposures as the reference frame in \texttt{ccmap}. \label{ccmapsol2}}
\end{figure}

\subsection{Creating the WFC3 Master Catalog}\label{wfc3master}
We take the resulting catalogs obtained after applying the procedure explained in Section~\ref{sec:wfc3phot} and group them by target cluster. The objective is to obtain a WFC3 master catalog which uses the astrometric information from all different exposures of the same target cluster. First, we arbitrarily choose one of the exposures to be the frame of reference to be mapped onto all other image frames. Then, we identify multiple star pairs on the individual exposures of the same target cluster in order to find a first-order initial coordinate transformation that sets every catalog's coordinate system in the chosen frame of reference. Each exposure was taken with an arbitrary rotation angle of the spacecraft, therefore we need a tool that can find, without any initial information, star pairs in catalogs that have arbitrarily shifted and rotated coordinates relative to one another. For this, we use the {\sc IRAF} task \texttt{xyxymatch} with its \textit{triangles} matching algorithm. This process defines a sample of a few dozen bright, non-saturated star matches in each frame catalog. These star matches serve as the input for the {\sc IRAF} task \texttt{ccmap} which calculates the initial coordinate transformation of each catalog into the frame of reference. The catalog coordinates are then transformed using the linear transformation within the {\sc IRAF} task \texttt{ccsetwcs}. Once every catalog of a given target cluster is, to first-order, set to the same frame of reference, we repeat the process with \texttt{xyxymatch}, which now is able to find significantly more matching stars in all frames using only a few pixels as matching tolerance. 

We use the augmented, full list of  star matches, which now contains several thousands of stars as input for \texttt{ccmap} and allow a full 6-parameter fit for the transformation solution. We show in Figure~\ref{ccmapsol1} one representative set of residuals for a case of 6 different exposures of the same target cluster. Again, this solution is applied using \texttt{ccsetwcs} to all catalogs. At this point we perform a new source cross-match between every catalog of a target cluster and calculate the average $X$ and $Y$ coordinate of every source\footnote{This is true provided there is a detection in every exposure, which is not always the case for stars near the chip-gaps and edges of individual frames, which are not always covered by the detectors given the variations in pointing and spacecraft roll angle.}. This average $XY$ catalog is used in the following as the frame of reference and we reiterate the procedure. We show in Figure~\ref{ccmapsol2} the new resulting set of residuals from the same cluster catalogs as in the previous Figure. Note that there is an additional transformation coming from the catalog that was before used as the frame of reference, and is now also available to map into the average $XY$ catalog. As can be seen from the residuals shown in Figure~\ref{ccmapsol2}, this method is able to correct for most of the lower-frequency systematics. However, a small sawtooth residual effect, caused by CTE effects \citep{bag15} is still visible and hard to remove entirely at this point for most catalogs. We perform a new and last source cross-match on all catalogs of a target cluster and recalculate the average $X$ and $Y$ coordinate of every source, which becomes the final WFC3 master coordinate catalog of a given target cluster. An important positive outcome is that part of the sawtooth effect is removed as it tends to cancel out when averaging over the different sub-exposure catalogs. We find this effect to be no more than $\sim\!0.03$ pixels, and its impact on our measurements is discussed in the subsequent analysis sections.

\subsection{Mapping the WFC3 Master Catalog into the ACS Coordinate Frame}\label{secwfc3mapping}
At the beginning of our WFC3-to-ACS mapping procedure, we construct the WFC3 stellar luminosity function (LF) of every cluster and use it to detect the horizontal branch (HB) luminosity, which can be identified in the LF as a local overdensity in the brighter end. We then use the approximate F606W magnitude of the HB of each cluster as given in \cite{dot10} and construct the LF in the range of two magnitudes brighter and fainter than the given HB F606W magnitude level. This way we find the peak of the HB optical LF, which we assume to be populated by the same group of stars as the peak of the LF from the WFC3 data. We select a small group of stars around the HB LF peak in both the WFC3 master catalog and the ACS catalog and use again the IRAF task \texttt{xyxymatch} with its \textit{triangles} matching algorithm to find star matches. This does not give immediate results every time, and further manual interaction with the code is needed in some clusters for correct matches to be found.~In particular, in some clusters we had to fine-tune the \textit{nmatch} and \textit{tolerance} parameters until correct matches were found.~Once we have a small list of star-matches from both catalogs, it becomes basically a matter of repeating a similar procedure as explained in the previous subsection. First we use \texttt{ccmap} and \texttt{ccsetwcs} to set, to a first-order, the WFC3 master catalog of a cluster into the ACS coordinate reference frame. We then find a much larger list of matching stars using \texttt{xyxymatch} and use the full 6-parameter fit to find the new transformation solution with \texttt{ccmap}. 

\begin{figure}
\centering
 \includegraphics[width=\linewidth,bb=30 40 530 410]{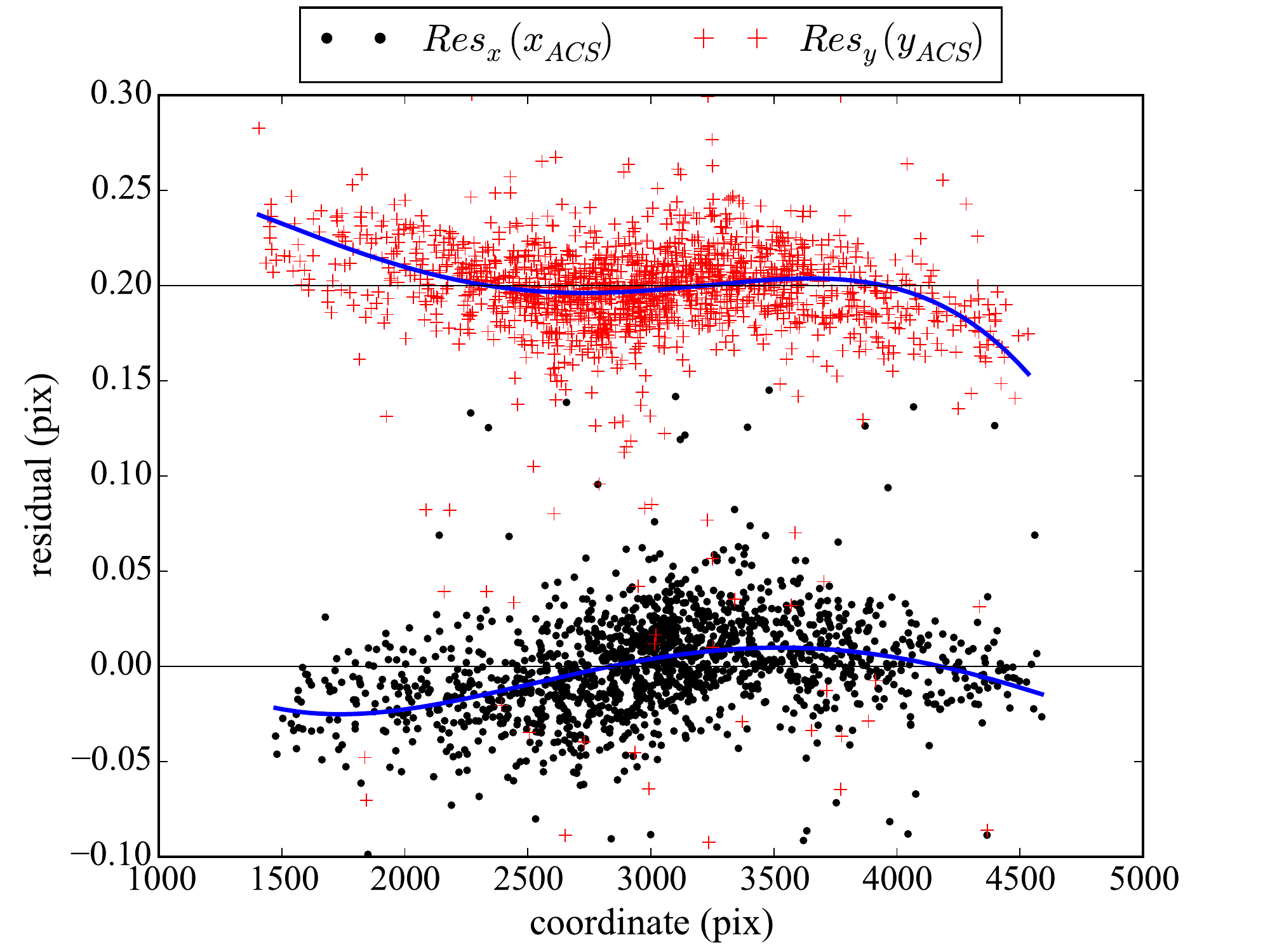} 
 \caption{Coordinate transformation residuals obtained with \texttt{ccmap}, which correspond to the mapping of the final NGC\,6717 WFC3 catalog into the ACS coordinate catalog. \tr{Black points and red crosses} show the residual in $X$ as a function of reference $X$ coordinate and the residual in $Y$ as a function of reference $Y$ coordinate, respectively. Note the residual dependence as a function of ACS pixel coordinates. This is caused by residual CTE systematics and the remaining CTE effects within the ACS catalog. The solid blue lines show a 4th-degree polynomial fit to the residuals.~The residuals in $Y$ have been shifted for clarity.  \label{ccmapsol3}}
\end{figure}

\begin{figure}
\centering
 \includegraphics[width=\linewidth,bb=30 40 530 410]{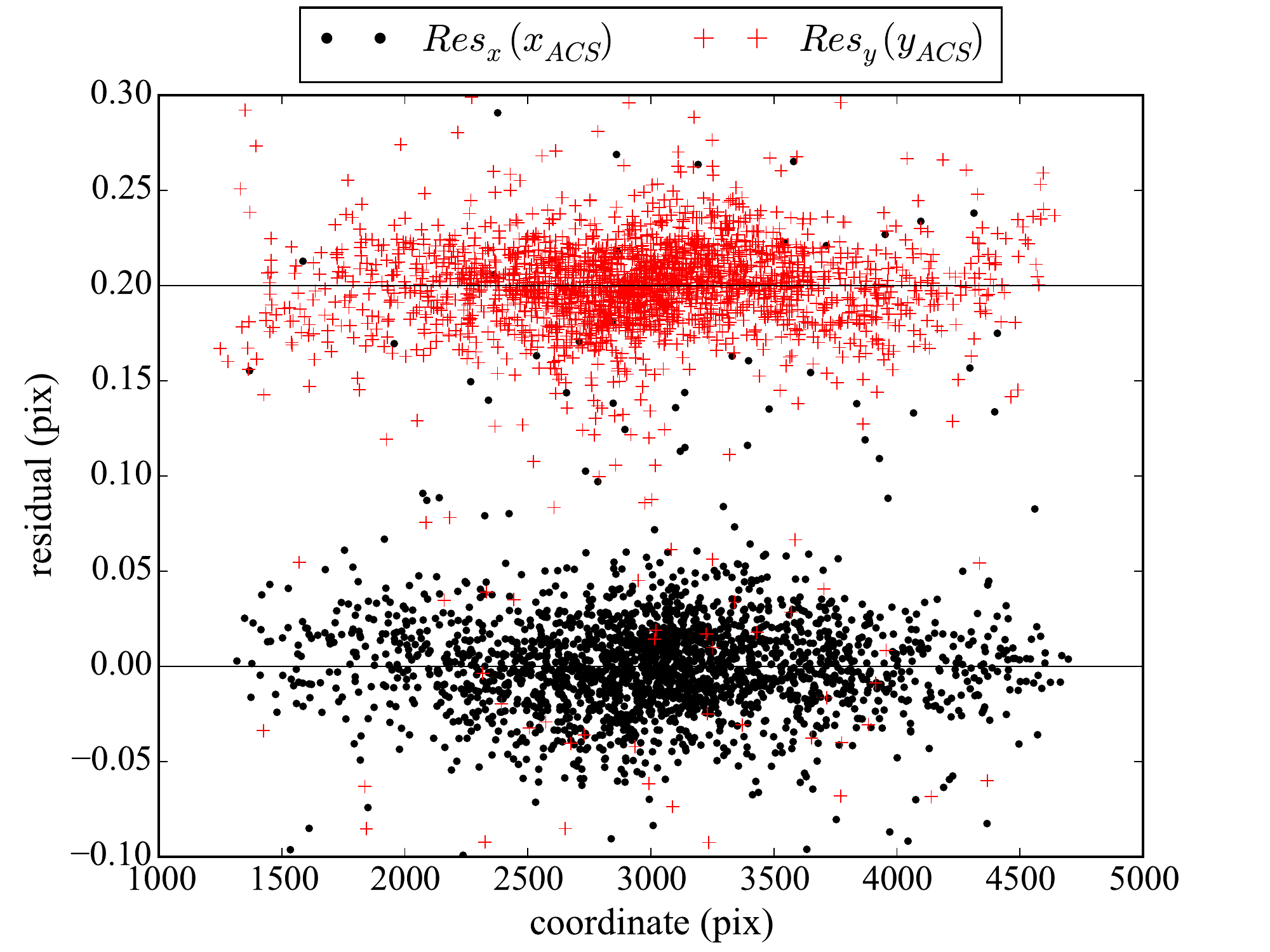} 
 \caption{Same as Fig.~\ref{ccmapsol3} after correcting for the polynomial fits from Fig.~\ref{ccmapsol3}.
 \label{ccmapsol4}}
\end{figure}

The solution is applied using \texttt{ccsetwcs} and then we perform a source cross-match using one pixel as matching tolerance. For the matched sources, we calculate now the preliminary proper motions and select the stars that have a proper motion within $1\sigma$ of the entire sample. These are assumed to have a high likelihood of being cluster members and, therefore, we use this new sample to recalculate the solution using the same procedure as explained above. If we therefore assume that we are transforming coordinates for cluster members only, hence not including non-members with high proper motions, then the best accuracy we can achieve is determined by the intrinsic instrumental uncertainty found when constructing the WFC3 master catalog. This is indeed the case, as can be seen in Figure~\ref{ccmapsol3}, where we confirm that the dispersion is about the same as in Figure~\ref{ccmapsol2}. We then fit a 4th-degree polynomial to the residuals and use it to correct for small lower-frequency systematics, which are still remaining in the transformation, likely caused by CTE effects that were not completely erased. This results in symmetric residuals independent of the physical coordinates on the camera detectors.~This is illustrated in Figure~\ref{ccmapsol4}, which demonstrates that we now have every WFC3 master catalog mapped into the ACS coordinate frame with an average uncertainty of $\sim\!0.02-0.03$\,pix.

\subsection{Second Source Detection in WFC3 frames}
The WFC3 master catalog coordinates are of the best accuracy we can obtain.~This is because we use the positions of detected sources in every F336W frame that is available, i.e.~we require a detection to be present in all F336W frames of a given target field.~However, this inevitably leads us to miss any star that happens to fall within the chip gap in any given exposure, even when this same star is detected in all other exposures of the same target field, not to mention failed detections due to source overlap with cosmic rays, bad pixels, and detector artifacts.~This is why we choose to perform a second detection procedure using now an $N\!-\!1$ detection condition, where $N$ corresponds to the number of available F336W sub-exposures for a given target cluster.~For most GCs, we have at least four available sub-exposures, except for NGC\,6341 and NGC\,6366, where only two F336W frames are available for photometry. For these two GCs we do not perform the second selection and coordinate measurement. For all the other GCs in our sample, we obtain new catalogs based on this new procedure, and redo the steps described in this section above in order to obtain new coordinate measurements, which will be of slightly lower accuracy than the ones previously measured with the full set of frames, as there is one less data point to measure the average position of each star in a given target field.~This new catalog is combined with the original catalog in the sense that only sources that were not included before will get added, hence not affecting the astrometry of starts that were already measured.~This method allows us to augment our proper-motion catalogs by $\sim\!30-50\%$.

\subsection{Proper-Motion Vector Diagrams}
We use the catalogs obtained above and perform a source cross-match against the ACS reference frame using a matching tolerance of one pixel. This procedure automatically removes from our final catalog all stars with proper motions larger than one pixel, which is acceptable given that those will most likely be foreground field stars, since the GC proper motions are expected to be significantly smaller. We now define the relative proper motion values as: 
\begin{eqnarray*}
\delta_x=X_{\rm WFC3}-X_{\rm ACS} &{\rm and}& \delta_y=Y_{\rm WFC3}-Y_{\rm ACS}
\end{eqnarray*}
and obtain relative proper motions in pixel units for every matched stellar source in the catalogs.~We construct vector point diagrams by plotting $\delta_x$ against $\delta_y$ for every cluster in our sample, which are shown in Figure~\ref{allCMD} for every sample GC.~The cluster members scatter around the zero value by construction, as our method defines the GC proper motion as the reference. Note also that the chosen frame of reference, i.e. the ACS Globular Cluster Survey catalogs, are constructed to have $x$ coordinates aligned with the right ascension (Ra) axis, and the $y$ coordinates aligned with the declination (Dec) axis, which of course serves nicely to obtain proper motion values in the standard axis of Ra and Dec.

\begin{figure*}
\centering
\includegraphics[width=0.8\linewidth,bb=30 -20 530 420]{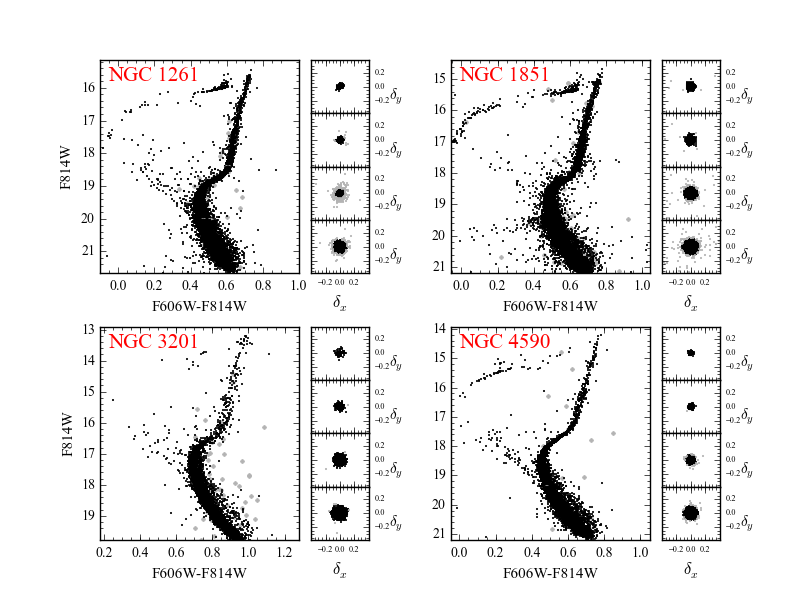} 
\includegraphics[width=0.8\linewidth,bb=30 0 530 400]{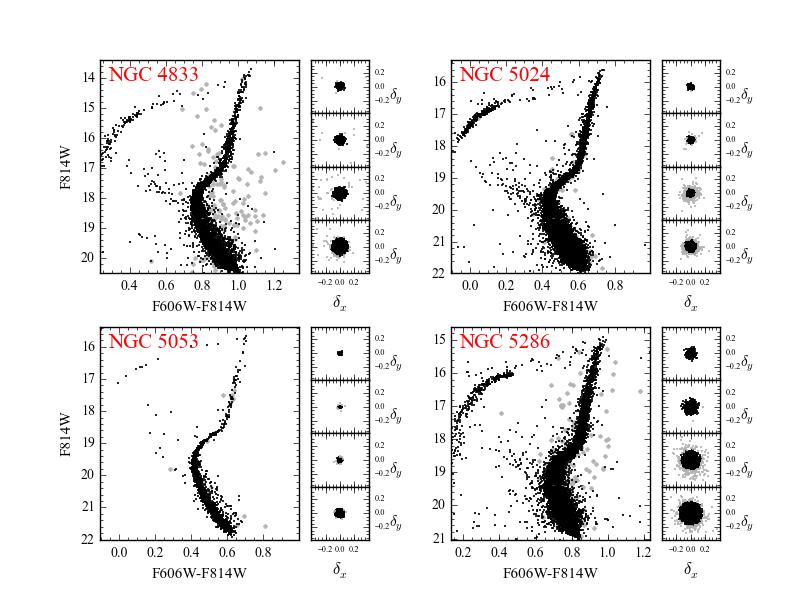} 
\caption{Color magnitude diagrams (CMDs) and relative proper motion distributions for our sample GCs. The vector point diagrams are shown in the smaller panels for four different magnitude ranges. Black points mark cluster member stars and grey points are non-members. }
\label{allCMD}
\end{figure*}

\begin{figure*}
\centering
\includegraphics[width=0.8\linewidth,bb=30 -20 530 420]{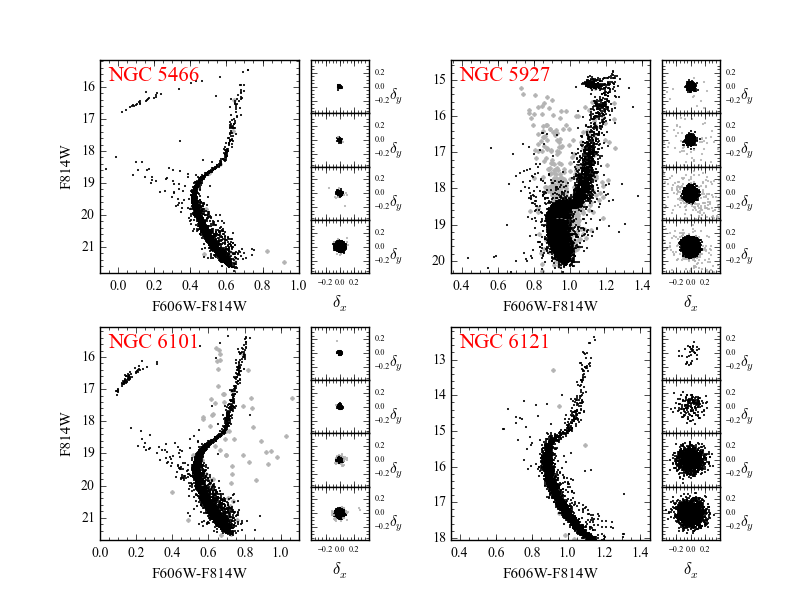} 
\includegraphics[width=0.8\linewidth,bb=30 0 530 400]{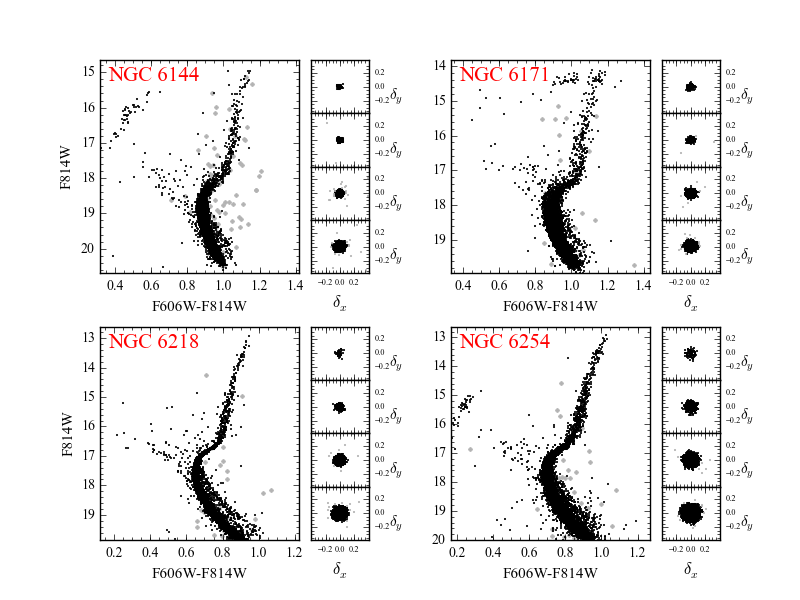} 
\contcaption{}
\end{figure*}

\begin{figure*}
\centering
\includegraphics[width=0.8\linewidth,bb=30 -20 530 420]{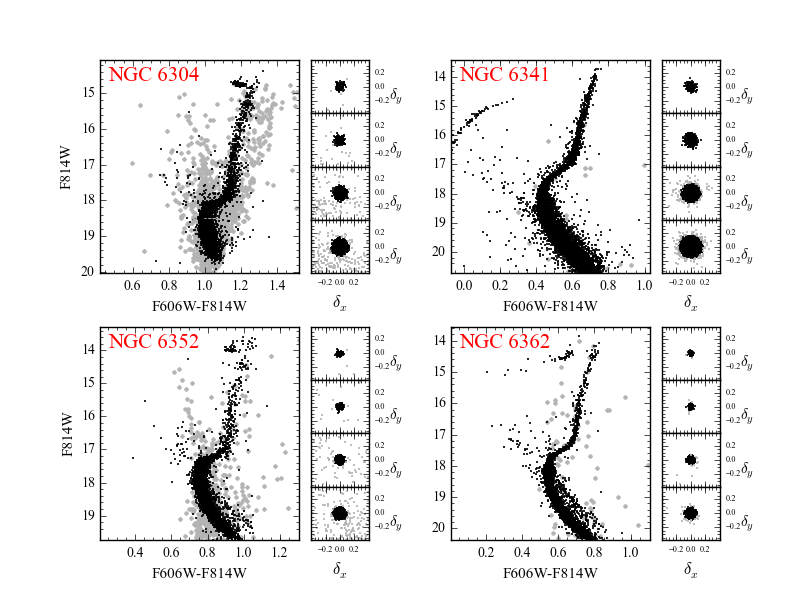} 
\includegraphics[width=0.8\linewidth,bb=30 0 530 400]{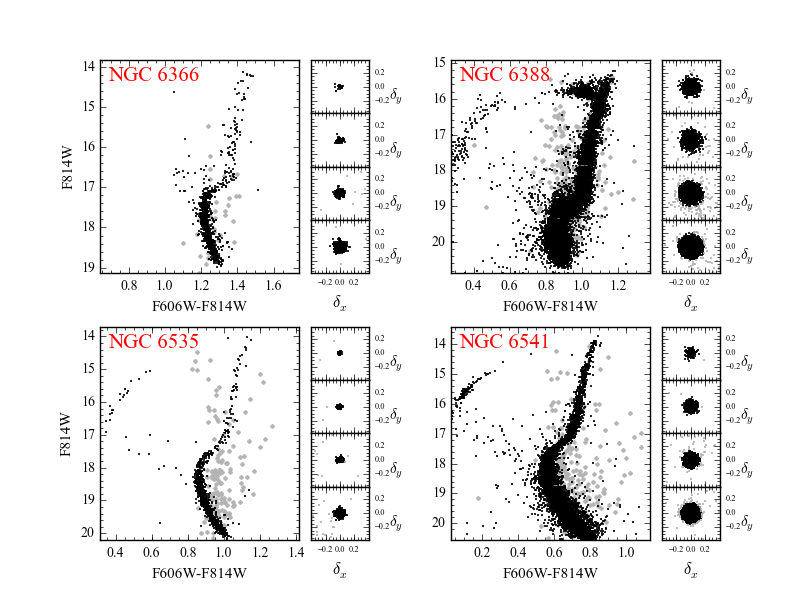} 
\contcaption{}
\end{figure*}

\begin{figure*}
\centering
\includegraphics[width=0.8\linewidth,bb=30 -20 530 420]{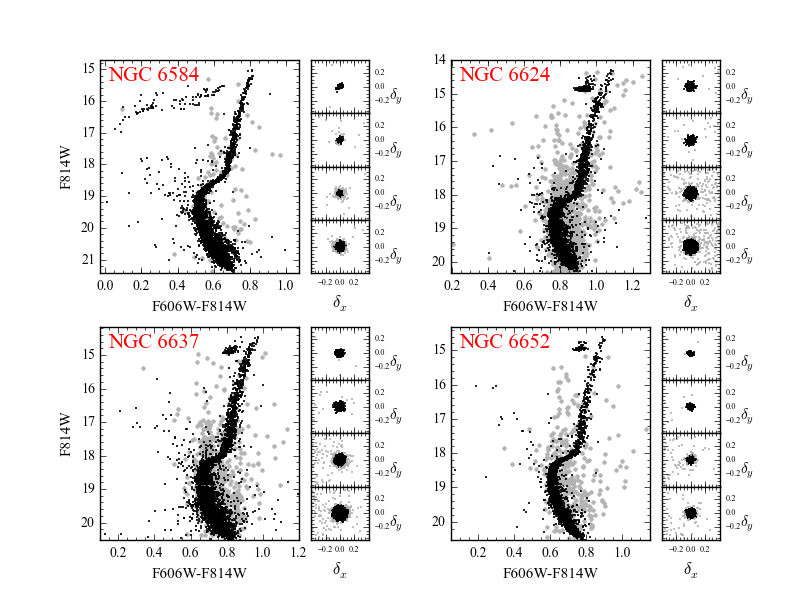} 
\includegraphics[width=0.8\linewidth,bb=30 0 530 400]{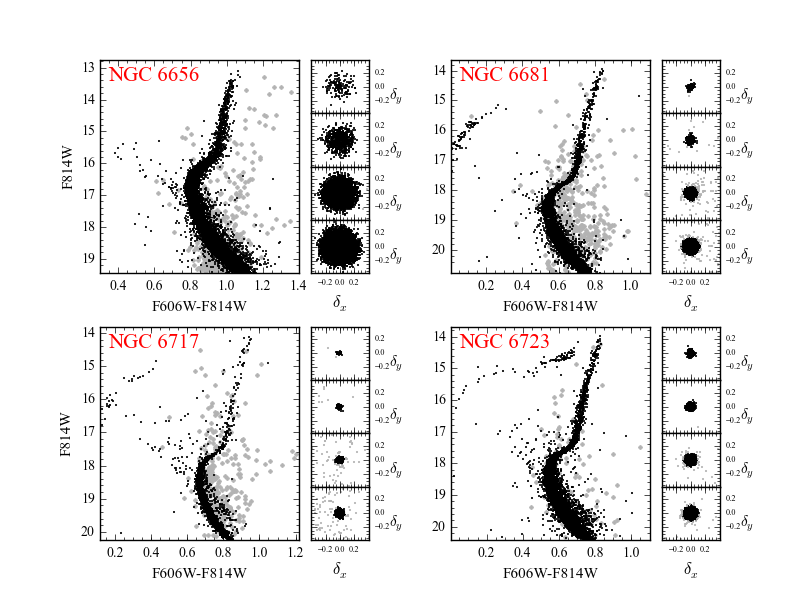} 
\contcaption{}
\end{figure*}

\begin{figure*}
\centering
\includegraphics[width=0.8\linewidth,bb=30 -20 530 420]{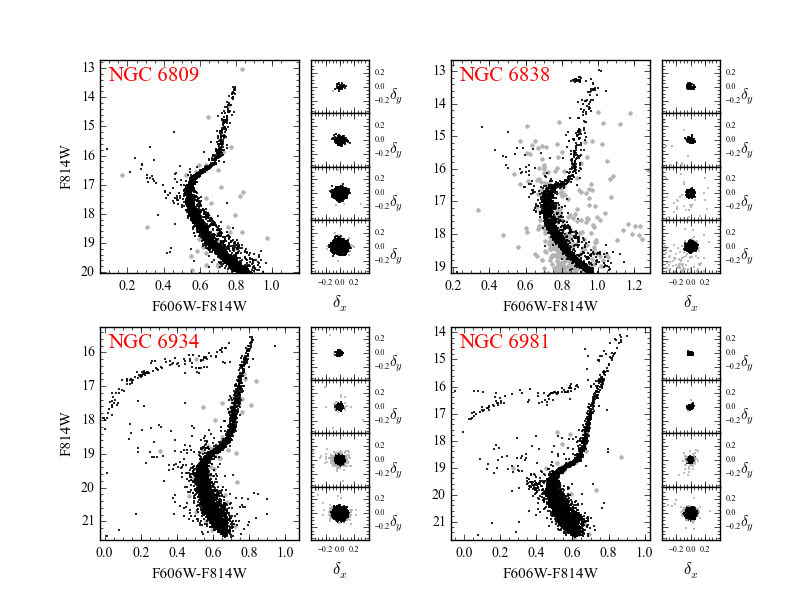} 
\includegraphics[width=0.8\linewidth,bb=30 0 530 400]{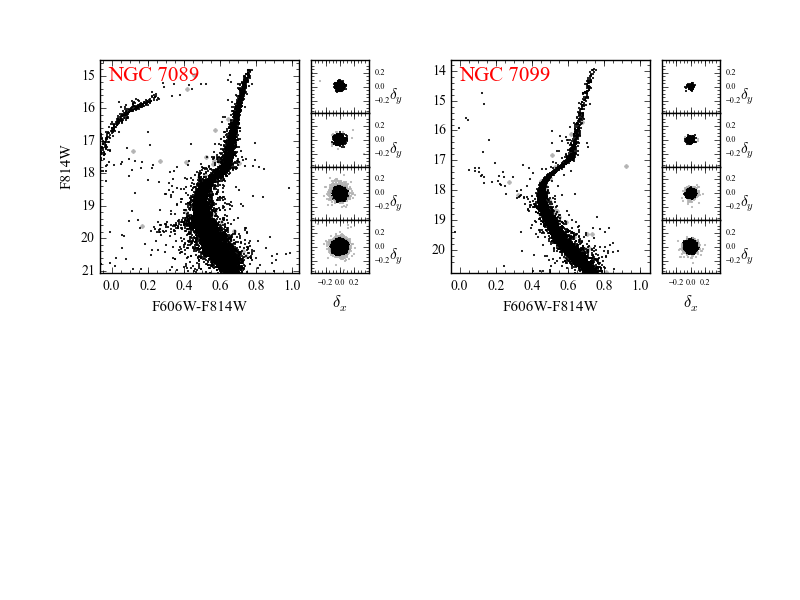} 
\contcaption{}
\end{figure*}

We use the proper motion information to select cluster members by studying the vector point diagram of stars at different luminosities. The underlying assumption is that more luminous (more massive) stars have, on average, different proper motions relative to less luminous stars. In particular, one expects from energy equipartition that more luminous stars have lower velocity dispersion than lower luminosity stars. Because of this, one needs to study the vector point diagram at different luminosity ranges. We divide each of the clusters catalog into four magnitude bins across the range of F606W magnitudes and construct their corresponding vector point diagrams (see Figure~\ref{allCMD}). For each subsample, we find an adequate selection criterion based on the observed dispersion of the vector point diagram. Because of the presence of very high proper motion stars, the measured dispersion can be unreliable, if measured directly, and therefore a more robust method must be applied in order to probe the true dispersion of the cluster stars. We choose to proceed as follows: 
\begin{enumerate}[{(1)}]
 \item We select stars with $R\!=\!\sqrt{\delta_x^2+\delta_y^2}\!<\!0.4$ pixels and calculate the standard deviation of $R$ for the particular subsample.  
 \item We remove stars with $R$ larger than $4.5\sigma$ and recalculate the standard deviation. 
 \item We iterate the process until the standard deviation changes by less than 4\%. This usually takes 3-4 iterations. 
 \item We classify stars as cluster members which have $R<\tr{6.5}\sigma$.
\end{enumerate}
\tr{It is important to note that when changing the threshold of the selection criterion, the stars inside the classical BSS region do not experience any significant fluctuations in their membership fraction (except for GCs with high contamination above the sub-giant branch). This reduces the possibility of accidentally removing BSSs with relative high proper motions values, which is very important since we do expect them to exist from dynamical interactions involving binaries}. Figure~\ref{allCMD} shows the CMDs of cluster member and non-member stars along with their vector point diagrams for every of our sample GCs.

The CMDs show that our proper-motion decontamination is very efficient at removing the scatter in the diagrams, whether it is due to noisy photometry or actual foreground/background star removal. Most notoriously, the proper-motion cleaning is able to successfully remove the stars usually contaminating the region above the sub-giant-branch (SGB). These objects are considered likely disk stars and are sometimes referred to as the "blue plume". This is very evidently shown for example in the CMDs of NGC\,5927, NGC\,6304 and NGC\,6652 (see Figure~\ref{allCMD}).~Our proper-motion cleaning method facilitates, therefore, a pure selection of a rich sample of BSSs in the classical CMD region, i.e.~brighter and bluer than the cluster main-sequence turn-off (MSTO) point, as well as above the SGB, where most previous studies usually failed to select BSSs, due to the significant field star contamination. 
\section{Analysis}
\label{txt:anls}

\subsection{Understanding the Errors}
It is important to test the posterior accuracy of our proper-motion measurements and their reliability when trying to associate these measurements to real dynamical information. The first instrumental error that we introduce to our measurements stems from the WFC3 master catalog, described in Section~\ref{wfc3master}. We calculate the average coordinates from multiple sub-exposures for each GC, and the final star coordinates in the WFC3 master catalogs are expected to carry an error caused partly by the motion of stars in the time span between the individual sub-exposures (up to $\sim$10 months for a few clusters). However, this is likely a negligible effect as we expect {\it (i)} these motions to be intrinsically less than\footnote{This assumes an average 0.3 mas/yr GC velocity dispersion \citep{wat15a}, considering the time span between exposures and the ACS pixel size.} 0.006 pix, and {\it (ii)} the minimum measurable proper-motion vector to be defined by much larger instrumental limitations \citep[i.e. the camera spatial resolution, the focus drift of each observation, etc.; see also][]{koz15}. 

\begin{figure}
\centering
\includegraphics[width=\linewidth,bb=30 0 530 410]{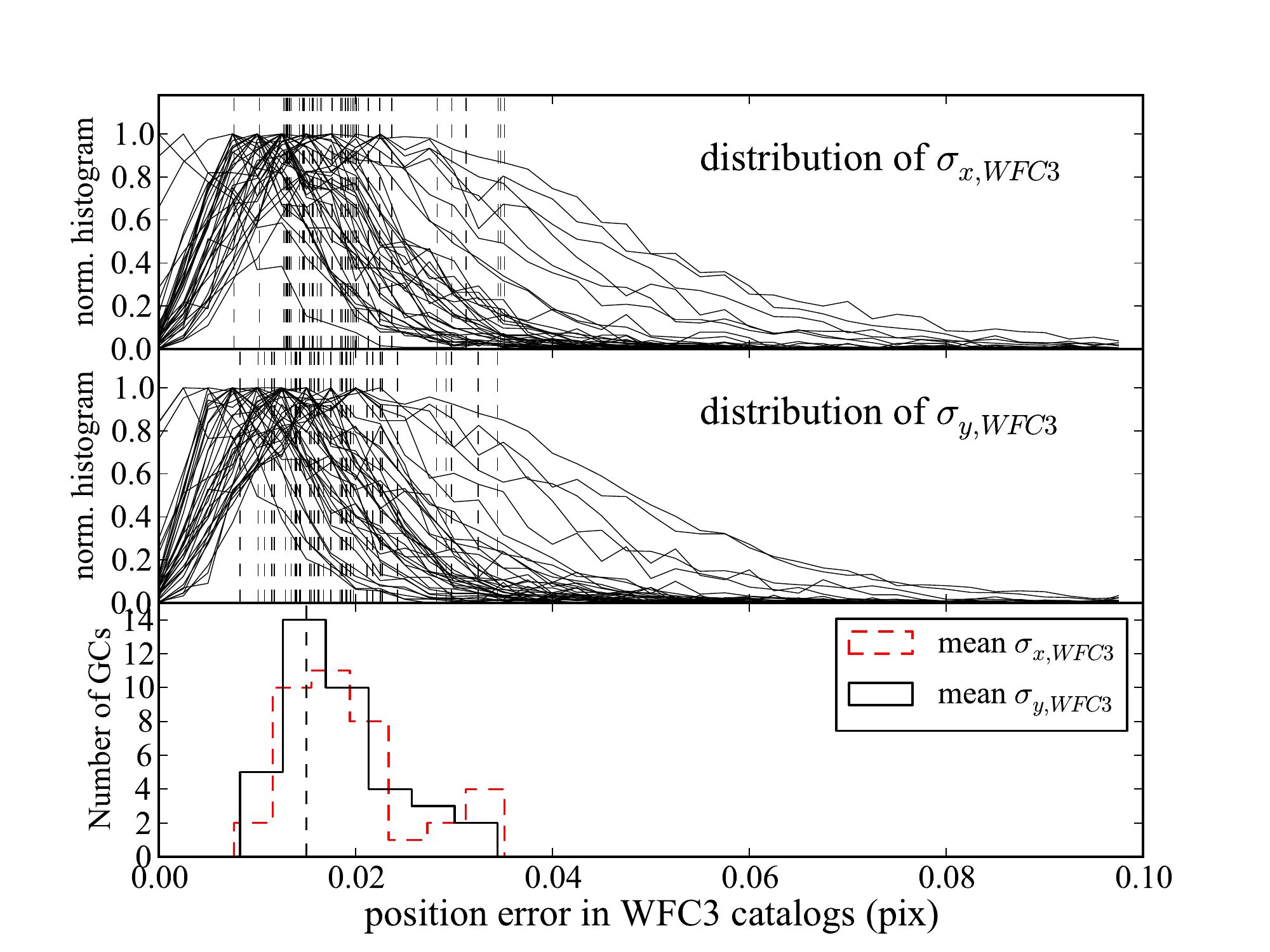} 
\caption{Normalized distributions of the standard deviation of different WFC3 coordinates of the cross-matched sources, when combining all different sub-exposure catalogs. The distributions are shown for the standard deviation of $X$ and $Y$ coordinates (top and middle panels, respectively) in every cluster of our sample, \trr{for stars down to one magnitude below the MSTO} (i.e.~each GC corresponds to a different solid line). The dashed vertical lines show the mean of the distribution for each GC catalog. The bottom panel shows the histogram of the mean $\sigma_x$ (red dashed line) and $\sigma_y$ (solid black line) for the GC sample. The vertical dashed line shows the value \trr{0.015} pix, which is the typical mean error in WFC3 coordinates.}
\label{wfc3errors}
\end{figure}

\begin{figure}
\centering
\includegraphics[width=\linewidth,bb=10 20 530 410]{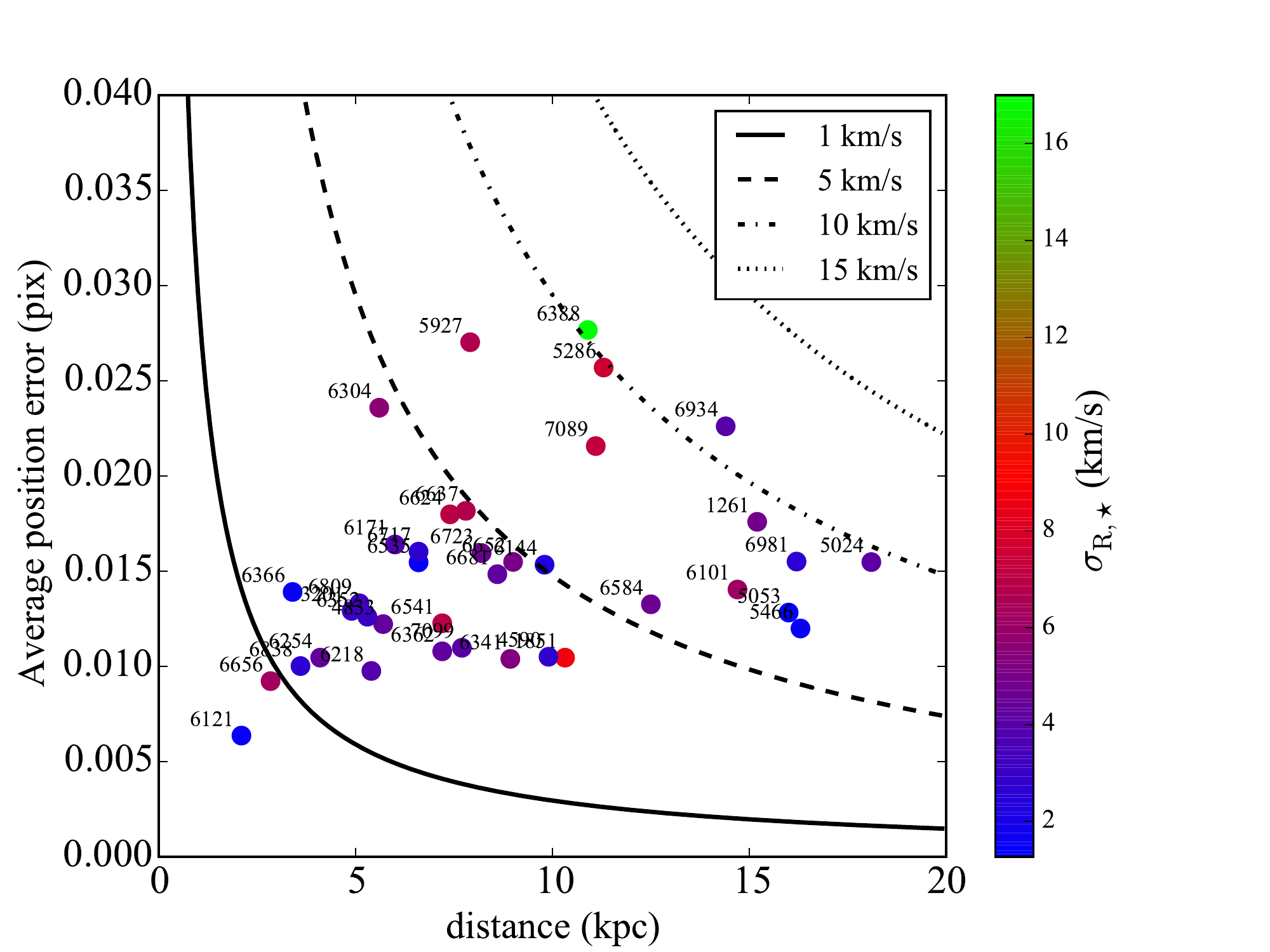} 
 \caption{\tr{Average position error (see Figure~\ref{wfc3errors}) in ACS pixel units for each of our target sample GCs as a function of the literature distance values from Section~\ref{comparison_watkins}. The color shading of the symbols encodes the expected central velocity dispersion (\trr{$\sigma_{\rm R,\star}$}) described in Section~\ref{comparison_watkins} for each GC. The black curves show the expected proper-motion dispersion value as a function of distance for stellar systems with four different central velocity dispersions. These lines represent the maximum allowed errors to robustly measure the corresponding central velocity dispersion value in a GC (see legend). We calculate these relations assuming a seven-year time baseline.}}
\label{testerrors}
\end{figure}

We show in Figure~\ref{wfc3errors} the positional error distributions in the $X$ and $Y$ coordinates of the WFC3 catalogs for all GCs in our sample. The errors are calculated by taking the standard deviation of the different position measurements for \trr{stars down to one magnitude fainter than the MSTO} in each cluster field. The figure shows that the measured coordinate uncertainty is on average about \trr{0.015} pixels for WFC3 coordinates, \tr{although we must note that the peak of the error distribution is a value \trr{$\sim\!20\!-\!30\%$ smaller}, but the long tail due to the more abundant faint main-sequence stars causes the average value to increase significantly. We still adopt the mean as a representative error to be as conservative as possible}. This uncertainty in the position of each star is the fundamental lower limit in our analysis and is inevitably carried over into the transformed coordinates in the mapped catalogs described in Section~\ref{secwfc3mapping}. Considering the difference in pixel size between ACS/WFC (0.05\arcsec/pix) and WFC3/UVIS (0.04\arcsec/pix), we expect the measured positions and proper motions in the ACS frame coordinates to have, in general, an inherent error of about \trr{0.01-0.02} pixels. This uncertainty has an impact on our proper motion measurements depending on the cluster distance and the projected motions of stars, i.e. for more distant clusters the mapping error becomes increasingly more dominant compared to the smaller stellar proper motions. This is best seen in Figure~\ref{testerrors} where the \tr{black lines show the expected proper motion dispersion as measured in ACS pixel units, as a function of distance, assuming four different velocity dispersions and a time baseline of seven years. We also show explicitly the average position error (see Figure~\ref{wfc3errors}) in ACS pixel units for each of our target sample GCs as a function of distance. The color shading of the symbols encodes the expected velocity dispersion value (see Section~\ref{comparison_watkins}). The black lines hence illustrate the maximum position error allowed to robustly measure the corresponding velocity dispersion value in a GC. For example, the GCs that in the diagram lie above their corresponding maximum error, for a given velocity dispersion, would most likely show a larger proper motion dispersion than what is expected}. This can be directly tested by estimating the central velocity dispersion for our cluster sample \tr{based on the measured proper motion dispersions.}

\subsection{Comparison with Literature Values}\label{comparison_watkins}
\begin{figure*}
\centering
\includegraphics[width=0.49\linewidth,bb=10 20 530 410]{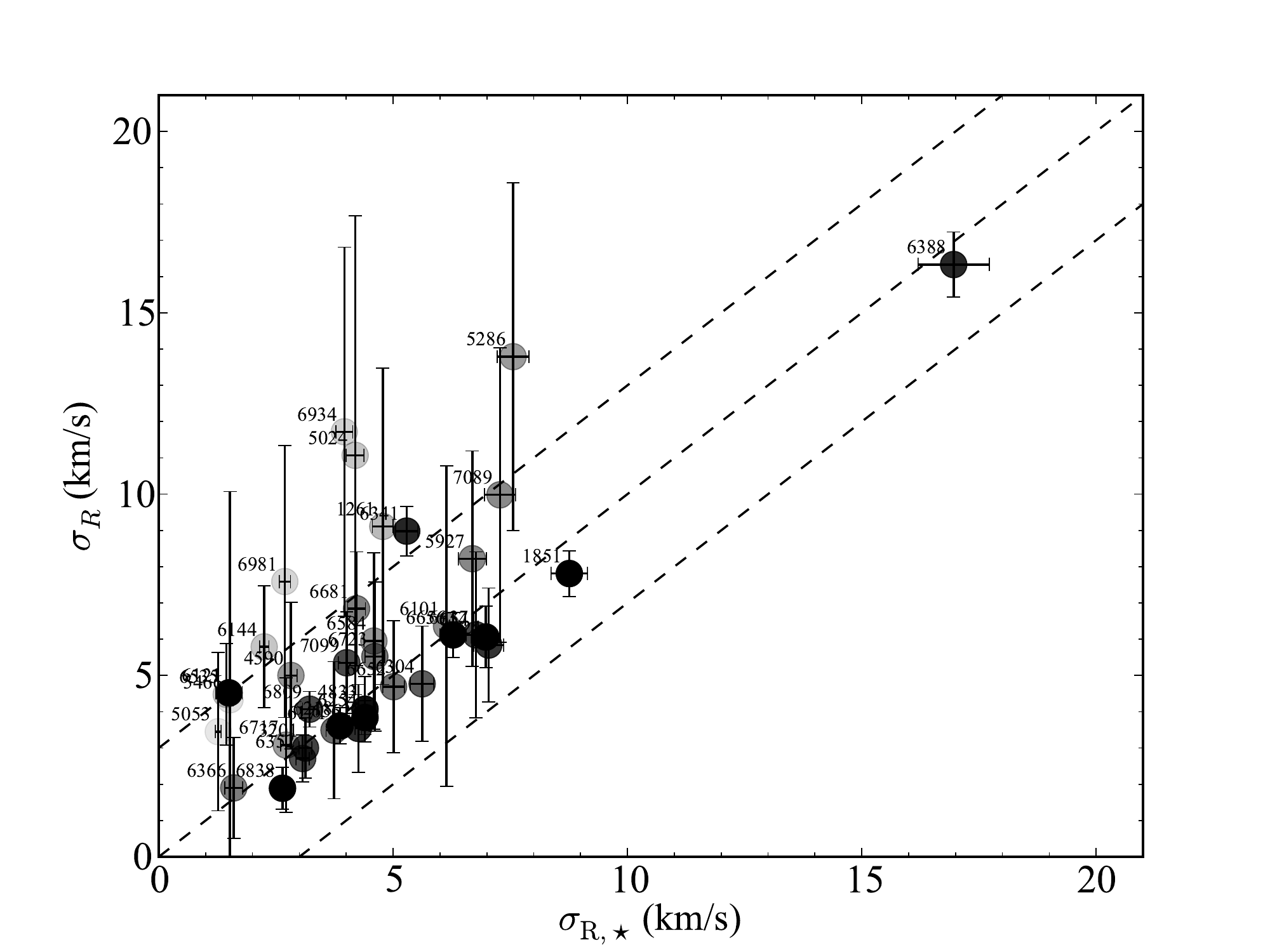} 
\includegraphics[width=0.49\linewidth,bb=10 20 530 410]{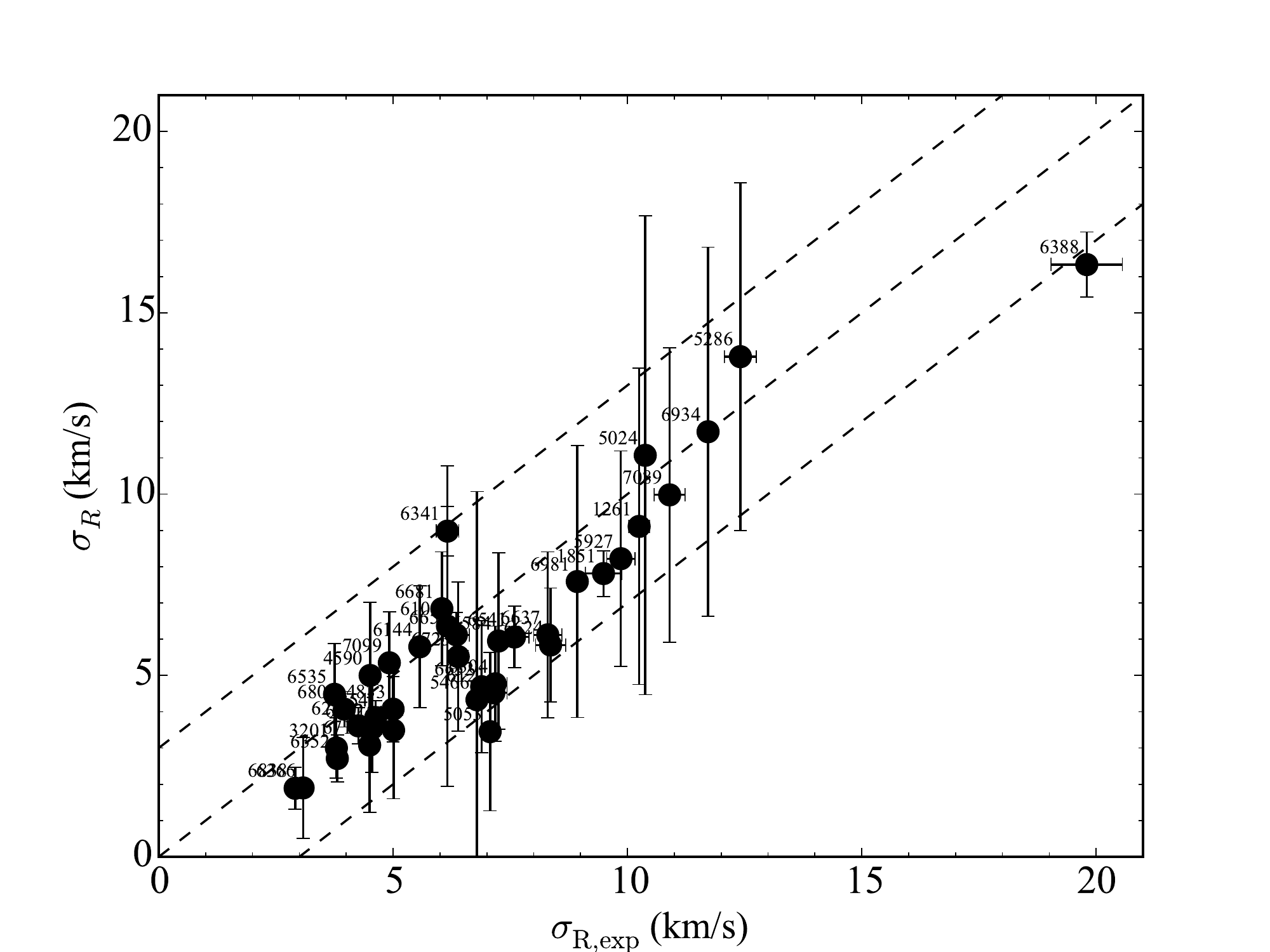} 
\caption{\trr{\textit{Left}:} Central velocity dispersion estimates from our proper-motion analysis ($\sigma_{\rm R}$) as a function of the expected velocity dispersion values \trr{($\sigma_{\rm R,\star}$)} from photometric considerations (see Section~\ref{comparison_watkins}). The error bars are calculated from the individual position errors of each star used in the dispersion estimate, while the errors of the expected values are calculated assuming a 0.1 mag uncertainty in the GC absolute magnitude. The symbol gray shading is scaled to the significance with which the expected velocity dispersion (\trr{$\sigma_{\rm R,\star}$}) would be measured for each GC given its average position error (see Figure~\ref{wfc3errors}). Therefore the lighter points with large error bars and systematically above the 1-to-1 relation correspond to GCs that have large mean position errors and are located at a relatively large distance, i.e. in the upper right region of Figure~\ref{testerrors}. \trr{\textit{Right}: Central velocity dispersion estimates from our proper-motion analysis ($\sigma_{\rm R}$) as a function of the expected \textit{observed} velocity dispersion values \trr{($\sigma_{\rm R,exp}$)}. The x-axis values are calculated as the combination of the intrinsic velocity dispersion ($\sigma_{\rm R,\star}$) values plus the dispersion broadening coming from the individual GC proper motion errors. }The dashed lines show the one-to-one relation as well as the $\pm$3 km/s region which is representative of the scatter expected from the error bars shown.}
\label{expected_sigma}
\end{figure*}

\tr{For the GCs in our sample we plot \trr{in the left panel of} Figure~\ref{expected_sigma} the dispersion of $R$ (\trr{$\sigma_R$}) from all member stars down to luminosities one magnitude fainter than the MSTO within \tr{10}$\arcsec$ from the cluster center (representing the 2-D central velocity dispersion $\sigma_R=\sigma_0\sqrt{2/3}$) versus the expected $\sigma_{\rm R,\star}$ \trr{based on the GC luminosity} via}
\begin{equation*}
\sigma_{\rm R, \star} = \sqrt{\frac{2{\cal M}_\star G}{3\beta r_h}},
\end{equation*}
where ${\cal M}_\star$ is the GC stellar mass\footnote{\tr{The stellar masses were computed from the $V$-band mass-to-light ratios from the SSP model predictions of \cite{bc03}, which were computed for each corresponding GC metallicity assuming a stellar population age of 12 Gyr. The absolute $V$-band magnitudes and metallicities were taken from \cite{har10}.}}, \tr{$r_h$ the GC half-light radius, and $\beta$ the dynamical scaling parameter, for which we adopt $\beta=12$ which is generally used for compact stellar systems \citep[see][for details]{cap06}.  }

\begin{figure}
\center
\includegraphics[width=\linewidth,bb=10 20 530 410]{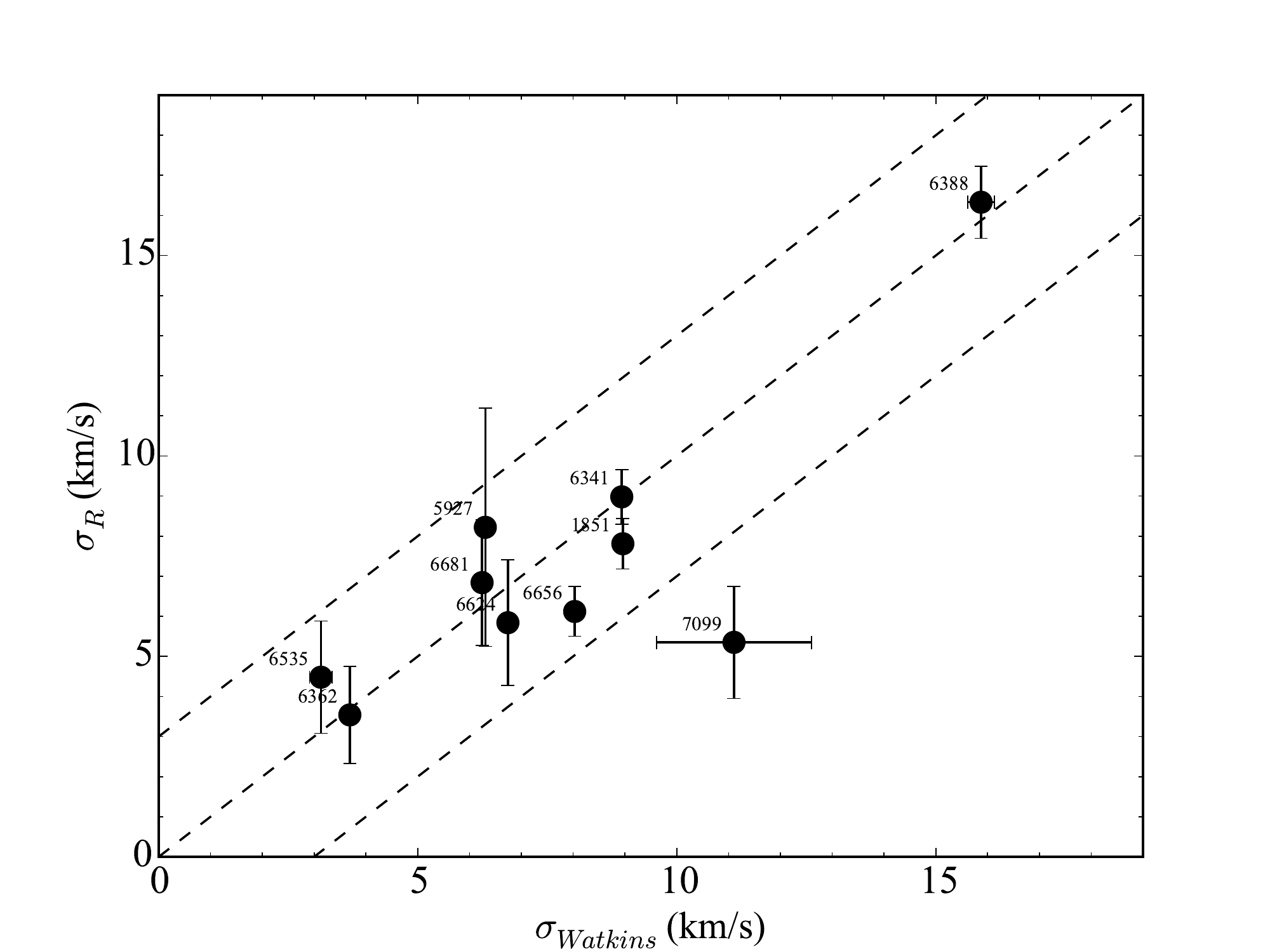} 
\caption{Central velocity dispersion estimates from our proper motion data compared to the ones from \protect\cite{wat15a}. \tr{Their proper motion dispersion values are converted into velocities using the distances from \protect\cite{wat15b}}.}
\label{watkins}
\end{figure}

We transform the projected motions to velocities assuming a seven year time span and the cluster distances given by \cite{wat15b}, or by \cite{har10} for those clusters that are not in the former catalog. We find that most of our estimates show a general agreement with the \tr{expected} values. \tr{However, as it was predicted, for the cases where the position error from Figure~\ref{testerrors} becomes significantly larger than the expected proper motion dispersion, we obtain a systematic overestimation of the velocity dispersion, as it is clearly seen in the GCs with large error bars. The symbols gray shading are scaled to $\sigma_{\rm R,\star}$ in pixel units divided by the mean instrumental error for each GC, which basically corresponds to the significance with which the expected velocity dispersion ($\sigma_{\rm R, \star}$) would be measured for each GC given its average instrumental error. \trr{Therefore the lighter points show GCs in which the $\sigma_R$ values should be more affected by the errors. In the right panel of Figure~\ref{expected_sigma} we try to recover our measured $\sigma_R$, taking now into account how the instrumental errors would artificially broaden the intrinsic velocity dispersion. Hence, we sum in quadrature the average position measurement uncertainty ($\sigma_{\rm pos}$)  for each GC (see Fig.~\ref{wfc3errors}), and the $\sigma_{\rm R,\star}$ values. This way we try to mimic an expected {\it observed}\footnote{\trr{The reader should note that this quantity now differs from a realistic velocity dispersion, but is rather a direct attempt to reproduce our proper motion dispersion measurements, by taking into account how the errors would affect the real velocity dispersion.}} velocity dispersion $\sigma^2_{\rm R, exp}=\sigma^2_{\rm R, \star}+\sigma^2_{\rm pos}$. The plot shows good agreement, which means that our assumption that the overestimations in the $\sigma_R$ values are dominated by the measured positional errors seems valid.} This Figure nicely illustrates what considerations must be taken into account when interpreting our measurements, and what kind of accuracy we expect to achieve when comparing it to other similar measurements.} Subsequently, we now compare our estimates with the values from the recent study of \cite{wat15a}, who homogeneously derived central velocity dispersion values from HST proper motions of \tr{stars down to luminosities one magnitude below the MSTO (i.e. the same adopted limit in our study) in 22 GCs}, as part of the HSTPROMO collaboration \citep{bellini14}. We show the result of this comparison in Figure~\ref{watkins} for the ten clusters in common. All of these ten clusters \tr{populate a region in Figure~\ref{testerrors} where we do not expect our measurements to be severely affected by the instrumental errors}. Indeed, the Figure shows good agreement between both studies. The scatter is entirely consistent with individual GC measurement errors, which again confirms that we are correctly estimating our uncertainties. 
\tr{We note the case of NGC\,7099, which shows a moderate discrepancy. Its position in Figure~\ref{testerrors} (at $\sim\!7.5$ kpc distance and \trr{$\sim\!0.01$\,pix} average error) puts it slightly above its allowed maximum error for its expected central velocity dispersion ($\sigma_{R,\star}\!\approx\!4$ km/s). This implies that our estimated velocity dispersion should be slightly affected by the measurement uncertainties and needs to be considered an upper limit. Indeed our estimate is $\sim$1 km/s larger than the expected value. Figure~\ref{watkins} then suggests that the value from \cite{wat15a} is overestimated by at least 6 km/s. This is actually consistent with their results (see their Figure 13), which shows their estimate to be $\sim$7 km/s larger than previous studies. It is worth mentioning that NGC\,7099 is actually the smallest data set from the entire cluster sample in their study, and hence the polynomial fit to their dispersion profile is the poorest out of their entire cluster sample. Considering the above, we conclude that a comparison between both measurements in NGC\,7099 has to be taken with care. Nevertheless, the overall agreement} with the \cite{wat15a} study illustrates that accurate proper motions can be reproduced by different data analysis methods to a very high degree and can be, therefore, used to derive absolute physical properties of the observed GCs, such as dynamical masses. Hence we conclude that, within our expected measurement uncertainties, our proper motion analysis can be used to produce clean BSS catalogs.

\begin{figure*}
\centering
 \includegraphics[width=0.8\linewidth]{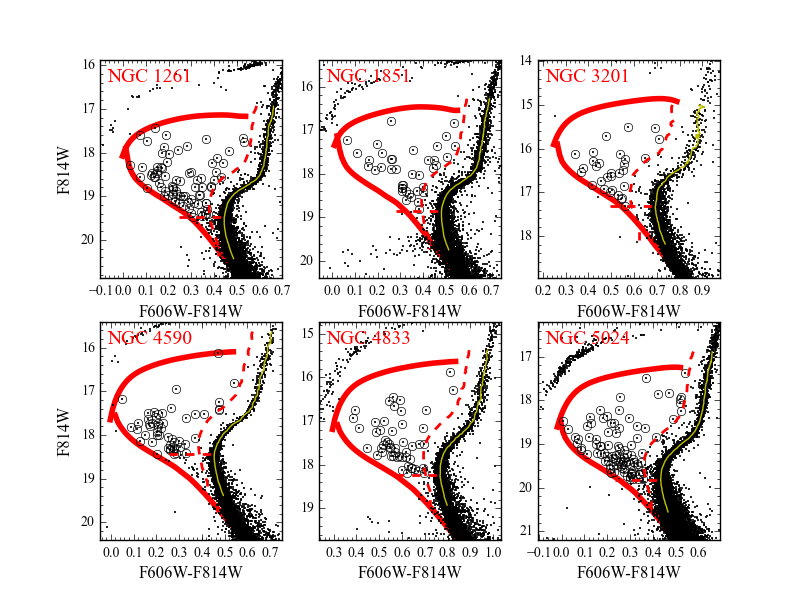} 
 \includegraphics[width=0.8\linewidth]{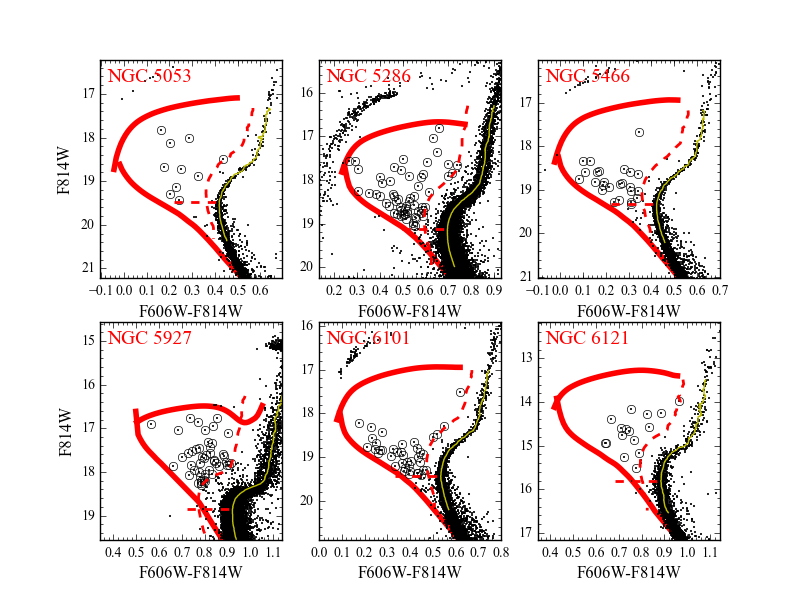} 
 \caption{Color magnitude diagrams of our globular cluster sample. The selected BSSs are labeled as circles. The red thick lines mark the isochrone-based limits for the BSS selection region. The yellow solid line shows the FL and the red dashed line shows the FL-based red limit for the BSS selection region. \tr{The horizontal red dashed line marks the empirical MSTO, as measured from the FL. } 
 \label{bss}}
\end{figure*}

\begin{figure*}
\centering
 \includegraphics[width=0.8\linewidth]{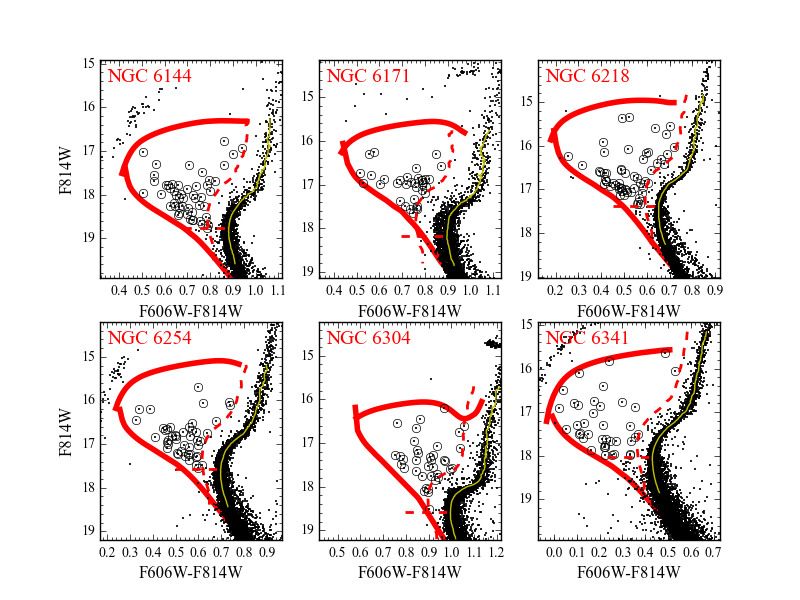} 
 \includegraphics[width=0.8\linewidth]{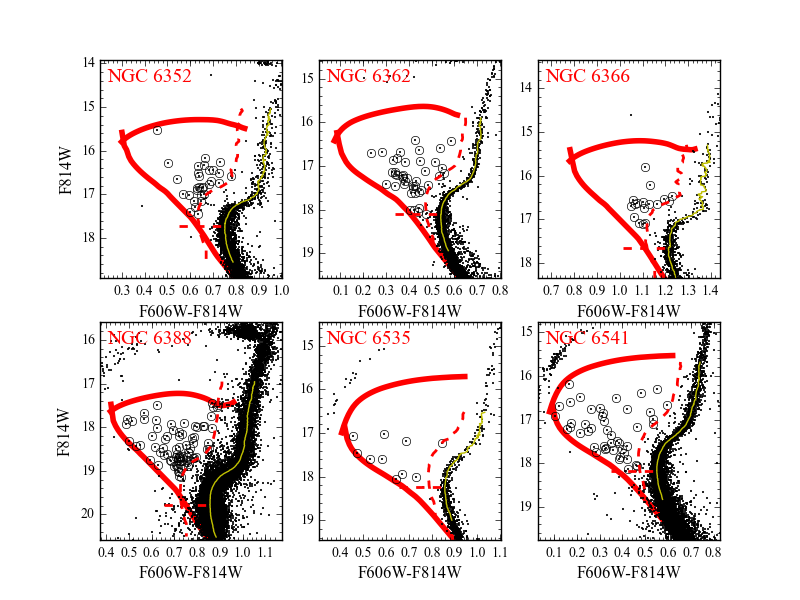} 
\contcaption{}
\end{figure*}

\begin{figure*}
\centering
 \includegraphics[width=0.8\linewidth]{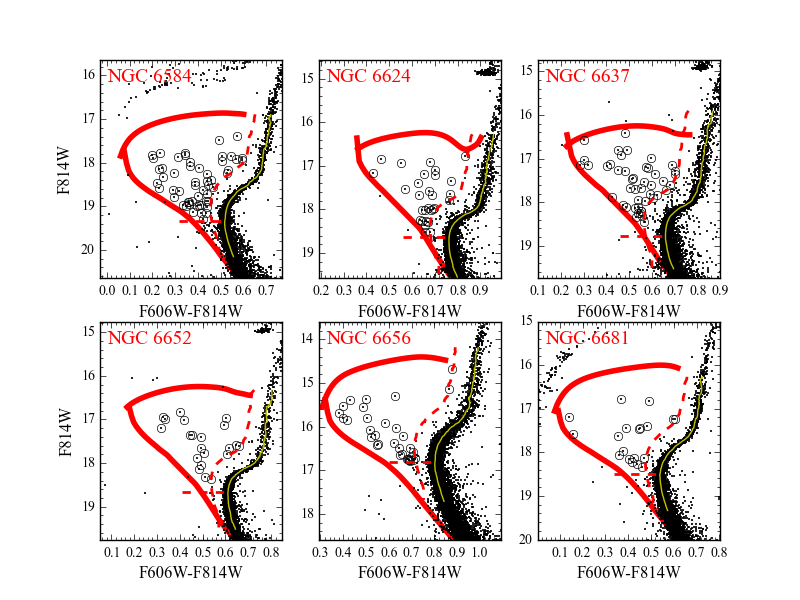} 
 \includegraphics[width=0.8\linewidth]{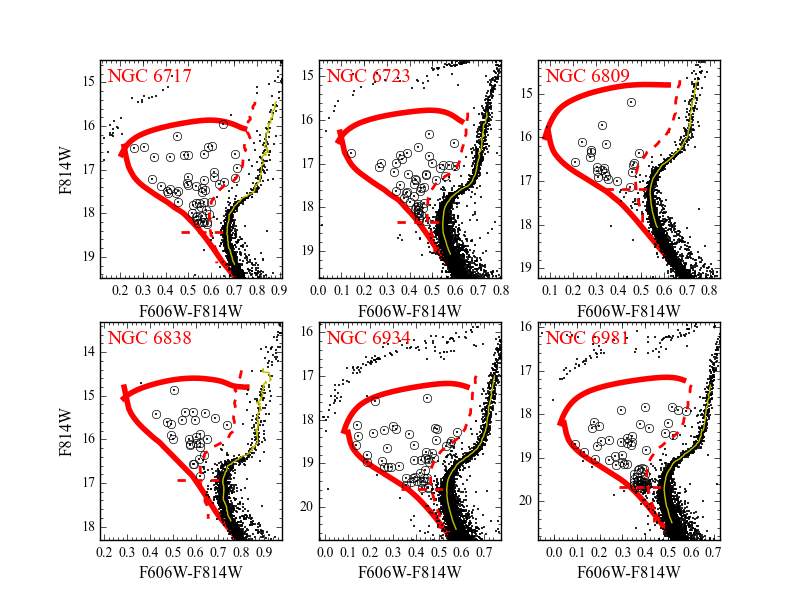} 
\contcaption{}
\end{figure*}

\begin{figure*}
\centering
 \includegraphics[width=0.85\linewidth, bb=0 190 612 432]{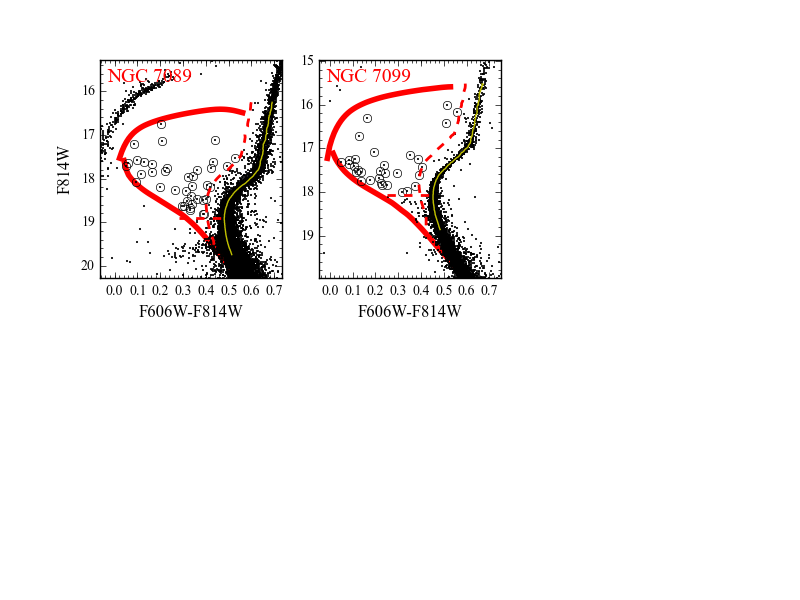} 
\contcaption{}
\end{figure*}

\subsection{Selection of BSS Candidates}
We now have stellar catalogs of our sample GCs that are free from fore- and background field stars, which may contaminate the BSS region in the CMDs between the sub-giant branch and the horizontal branch (see Figure~\ref{bss}).~This is mainly due to the young stellar populations of the Milky Way disk.~The next step of the BSS selection process is to homogeneously define a BSS selection function that can be uniformly applied to the CMD of every GC.

We compute the ridge line (RL) of GC stars along the main sequence (MS), sub-giant branch (SGB) and red-giant branch (RGB) directly from the data. We determine the spread in F606W-F814W color around the RL at each F814W magnitude and use it to define a red limit for BSS selection. We define stars within $4.5\sigma$ around the RL to be members of the MS, SGB or RGB, while stars falling $>\!4.5\sigma$ towards bluer colors are considered BSS candidates. This limits the selection of BSS candidates towards red colors. To set a limit towards bluer colors we then use Dartmouth isochrones \citep{dot08} with the GC  stellar population parameters found by \cite{dot10} or from the McMaster catalog \citep{har10} for GCs which are not listed in the former catalog. We use a 1-Gyr old isochrone at an $\alpha$ enhancement of 0.2\,dex for all GCs to define the bluest limit for our BSS selection. For most GCs, such an isochrone intersects in its SGB part with portions of the horizontal branch (HB) GC stellar population. We, therefore, use a copy of the 1-Gyr isochrone which is shifted 1\,mag to fainter luminosities to limit our BSS selection at the bright end. The combination of these three curve sections outlines a closed parameter space region in the F814W vs. F606W-F814W CMD and defines our BSS candidate selection function (see Figure~\ref{bss}). \tr{Additionally, we require for the selected BSSs to be brighter than the empirical MSTO magnitude, defined as the bluest point of the RL}.~\tr{We point out that previous works used similar HST data to derive their BSS catalogs \citep[e.g.][]{leigh11a}, while restricting their chosen BSS selection criteria to the classic BSS sub-population. Their definitions were using semi-empirically chosen polygons in the corresponding CMDs that were motivated by earlier studies \citep{ferr97, deM06, leigh07} and were avoiding the so-called yellow straggler star sub-populations \citep[e.g.][]{hess84, mccl85, mcga14}. Our definition is entirely data-based and includes the main BSS as well as the yellow straggler locus.} We note that this CMD region is also commonly contaminated by young disk stars -- a feature usually referred to as the blue plume. This is clearly seen for example in NGC\,5927, NGC\,6304, NGC\,6352, NGC\,6388, and NGC\,6624 as the high concentration of grey points above the SGB (see Fig.~\ref{allCMD}). Our proper motion decontamination shows that most GCs have at least some contamination in this region. The BSS selection presented here can, therefore, safely include stars in this region. \tr{This is actually a critical point for measurements such as the BSS fraction. In a recent paper, \cite{salinas16} showed that previous claims of M\,80 having a notably large BSS fraction were most likely affected by a severe contamination of such young disk stars and/or blends. Their results actually put the BSS fraction of M\,80 at a normal value for GCs.} \tr{We highlight the importance of the stars red of the classical BSS locus because possible interpretations for their existence may include: (i) the possibility of them being evolved BSSs, i.e.~older BSSs that are currently evolving through the Hertzsprung gap, and (ii) the possibility that they are recent collision products in the process of contracting back down to the ZAMS, i.e.~extremely young BSSs.} 

\subsection{Potential contaminants in the BSS sample}
\tr{We need to consider the possibility of remnant contaminants that passed our selection function. Foreground and background stars are effectively removed by the proper motion technique, therefore they can be safely considered non-existent in our sample. The possibility of star blends deserves some discussion. For example, random alignments between very hot stars, e.g. blue horizontal branch stars (BHBs), and cooler stars can easily mimic optical BSS colours. This is indeed very problematic for BSS selection with ground-based seeing-limited imaging data. For this reason, early HST studies \citep{ferr97} combined near-UV and optical imaging to exclude such contaminants. In purely optical CMDs, such blends between BHB+cooler stars (i.e.~MSTO, SGB, and RGB stars), can fall in the classical BSS and yellow straggler locus. Blends between MSTO and SGB/RGB stars preferentially populate the yellow straggler region \citep[see their Figure 8]{salinas16}. Since BSSs are truly hotter SEDs than MSTO/SGB/RGB blends, combining near-UV+ optical imaging will effectively identify such blends and select only genuine BSSs with the appropriate near-UV flux and near-UV+optical color. Blends between BHB+cooler stars are distinguished by their significantly bluer near-UV+optical colors than genuine BSSs. In our study, the fact that proper motions are derived from combined near-UV-optical data, already mitigates this issue. In particular, in the scenario of stellar blends with different temperature components, the measured centroids can easily be influenced differently when seen in the near-UV or in the optical. Given the approximate FWHM size of ACS ($\sim\!2\!-\!3$ pixels), any chance alignment of stars at smaller distances would either get rejected by our initial QFIT threshold filtering (due to the imperfect PSF fitting in such cases) or by the proper-motion selection itself, as such blends will likely be cataloged as high-proper motion stars (considering that our typical maximum proper motion is $\sim\!0.1\!-\!0.2$ pixels for selected members, see Figure~\ref{allCMD}). We still consider the possibility that blends might contaminate our BSS sample, and hence we have built F336W-F814W CMDs and directly inspected all BSS samples. We found a very small amount ($\sim\!10-20$ in our entire BSS library) of potential BHB/MSTO/SGB/RGB blends based on their position in these near-UV+optical CMDs and have flagged and removed them from the samples. Such low contamination fractions demonstrate the power of proper-motion based BSS selection functions and reassure the purity of our final BSS sample.}

\subsection{Proper Motion Cleaned BSS Catalogs}
The result of this work are proper motion cleaned, homogeneously selected, HST photometric catalogs of BSSs in the inner $\sim\!3\arcmin\!\times3\arcmin$ of \tr{38} Milky Way GCs \tr{(see Table~\ref{tab:sample})}. In Table~\ref{tab:n1261}, we provide the location, photometry, and proper motions for the first 30 BSSs candidates in NGC\,1261.~For completeness, the proper motion values have been converted to mas/yr units, assuming a 7 year baseline and the ACS/WFC pixel size (0.05\arcsec/pix).~The full list of catalogs for each of our sample GCs is available electronically and can also be obtained via this web link\footnote{http://www.astro.puc.cl/$\sim$msimunov/research.html}.  

\begin{table*}
\caption{Properties of the target GC sample. Columns 1-7 were taken from \protect\cite{har10}. Column 8 is the calculated expected central velocity dispersion described in Section~\ref{comparison_watkins}.}
\label{tab:sample}
\begin{tabular}{llccrcccrrcc}
\hline
   NGC    & Name   &  RA(J2000) & DEC(J2000)  &[Fe/H]&$M_{\rm V}$& $r_h$ &$\sigma_{\rm R,\star}$  \\
              &              &  [hr:min:sec]  & [deg:min:sec]  &  [dex]  & [mag]    & [arcmin]  & [km/s]  \\
\hline

  NGC\,1261&         & 03:12:16.21&$-$55:12:58.4 &  $-$1.27& $-$7.80& 0.68& $ 4.78 \pm0.23$\\
 NGC\,1851&         & 05:14:06.76&$-$40:02:47.6 &  $-$1.18&  $-$8.33& 0.51&  $8.76 \pm0.41 $\\
 NGC\,3201&         & 10:17:36.82&$-$46:24:44.9 &  $-$1.59&  $-$7.45& 3.10&  $3.13 \pm0.15 $\\
 NGC\,4590&M\,68& 12:39:27.98&$-$26:44:38.6 &  $-$2.23&  $-$7.37& 1.51&  $2.82 \pm0.13 $\\
 NGC\,4833&         & 12:59:33.92&$-$70:52:35.4 &  $-$1.85&  $-$8.17& 2.41&  $4.40 \pm0.21 $\\
 NGC\,5024&M\,53& 13:12:55.25&$+$18:10:05.4&  $-$2.10&  $-$8.71& 1.31&  $4.19 \pm0.20$ \\
 NGC\,5053&         & 13:16:27.09&$+$17:42:00.9&  $-$2.27&  $-$6.76& 2.61&  $1.27 \pm0.06$ \\
 NGC\,5286&         & 13:46:26.81&$-$51:22:27.3 &  $-$1.69&  $-$8.74& 0.73&  $7.56 \pm0.36$ \\
 NGC\,5466&         & 14:05:27.29&$+$28:32:04.0&  $-$1.98&  $-$6.98& 2.30&  $1.52 \pm0.07$ \\
 NGC\,5927&         & 15:28:00.69&$-$50:40:22.9 &  $-$0.49&  $-$7.81& 1.10&  $6.69 \pm0.32 $\\
 NGC\,6121& M\,4 & 16:23:35.22&$-$26:31:32.7 &  $-$1.16&  $-$7.19& 4.33&  $1.50 \pm0.07 $\\
 NGC\,6101&         & 16:25:48.12&$-$72:12:07.9 &  $-$1.98&  $-$6.94& 1.05&  $6.14 \pm0.29 $\\
 NGC\,6144&         & 16:27:13.86&$-$26:01:24.6 &  $-$1.76&  $-$6.85& 1.63&  $2.25 \pm0.11 $\\
 NGC\,6171&M\,107&16:32:31.86&$-$13:03:13.6&  $-$1.02&  $-$7.12& 1.73&  $3.74 \pm0.18 $\\
 NGC\,6218&M\,12& 16:47:14.18&$-$01:56:54.7 &  $-$1.37&  $-$7.31& 1.77&  $3.87 \pm0.18 $\\
 NGC\,6254&M\,10& 16:57:09.05&$-$04:06:01.1 &  $-$1.56&  $-$7.48& 1.95&  $4.40 \pm0.21 $\\
 NGC\,6304&         & 17:14:32.25&$-$29:27:43.3 &  $-$0.45&  $-$7.30& 1.42&  $5.62 \pm0.26 $\\
 NGC\,6341&M\,92& 17:17:07.39&$+$43:08:09.4&  $-$2.31&  $-$8.21& 1.02&  $5.29 \pm0.25$ \\
 NGC\,6352&         & 17:25:29.11&$-$48:25:19.8 &  $-$0.64&  $-$6.47&2.05 &  $ 3.07 \pm0.14$ \\
 NGC\,6366&         & 17:27:44.24&$-$05:04:47.5 &  $-$0.59&  $-$5.74& 2.92&  $1.60 \pm0.08 $\\
 NGC\,6362&         & 17:31:54.99&$-$67:02:54.0 &  $-$0.99&  $-$6.95& 2.05&  $4.26 \pm0.20 $\\
 NGC\,6388&         & 17:36:17.23&$-$44:44:07.8 &  $-$0.55&  $-$9.41& 0.52&  $16.96\pm0.8 $ \\
 NGC\,6535&         & 18:03:50.51&$-$00:17:51.5 &  $-$1.79&  $-$4.75& 0.85&  $1.44 \pm0.07 $\\
 NGC\,6541&         & 18:08:02.36&$-$43:42:53.6 &  $-$1.81&  $-$8.52& 1.06&  $6.98 \pm0.33 $\\
 NGC\,6584&         & 18:18:37.60&$-$52:12:56.8 &  $-$1.50&  $-$7.69& 0.73&  $4.59 \pm0.22 $\\
 NGC\,6624&         & 18:23:40.51&$-$30:21:39.7 &  $-$0.44&  $-$7.49& 0.82&  $7.04 \pm0.33 $\\
 NGC\,6637&M\,69& 18:31:23.10&$-$32:20:53.1 &  $-$0.64&  $-$7.64& 0.84&  $6.77 \pm0.32 $\\
 NGC\,6652&         & 18:35:45.63&$-$32:59:26.6 &  $-$0.81&  $-$6.66& 0.48&  $5.01 \pm0.24 $\\
 NGC\,6656&M\,22& 18:36:23.94&$-$23:54:17.1 &  $-$1.70&  $-$8.50& 3.36&  $6.28 \pm0.30 $\\
 NGC\,6681&M\,70& 18:43:12.76&$-$32:17:31.6 &  $-$1.62&  $-$7.12& 0.71&  $4.22 \pm0.20 $\\
 NGC\,6717&Pal\,9& 18:55:06.04&$-$22:42:05.3 &  $-$1.26&  $-$5.66& 0.68&  $2.72 \pm0.13 $\\
 NGC\,6723&         & 18:59:33.15&$-$36:37:56.1 &  $-$1.10&  $-$7.83& 1.53&  $4.61 \pm0.22 $\\
 NGC\,6809&M\,55& 19:39:59.71&$-$30:57:53.1 &  $-$1.94&  $-$7.57& 2.83&  $3.22 \pm0.15 $\\
 NGC\,6838&M\,71& 19:53:46.49&$+$18:46:45.1&  $-$0.78&  $-$5.61& 1.67&  $2.64 \pm0.12$ \\
 NGC\,6934&         & 20:34:11.37&$+$07:24:16.1&  $-$1.47&  $-$7.45&0.69 &  $ 3.96 \pm0.19$ \\
 NGC\,6981&M\,72& 20:53:27.70&$-$12:32:14.3 &  $-$1.42&  $-$7.04& 0.93&  $2.69 \pm0.13 $\\
 NGC\,7089&M\,2  & 21:33:27.02&$-$00:49:23.7 &  $-$1.65&  $-$9.03& 1.06&  $7.28 \pm0.34$ \\
 NGC\,7099&M\,30& 21:40:22.12&$-$23:10:47.5 &  $-$2.27&  $-$7.45& 1.03&  $4.01 \pm0.19 $\\

\hline
\end{tabular}
\end{table*}

\begin{table*}
\caption{Properties of selected Blue Straggler Stars in NGC\,1261. Columns 1-5 are taken from the original ACS catalogs from \protect\cite{sara07}. Columns 6 and 7 correspond to the proper motion values in sky coordinates, taking into account the cos(Dec) correction. They have been converted to mas/yr units, assuming a 7-year baseline and the ACS/WFC pixel size (0.05\arcsec/pix).}
\label{tab:n1261}
\begin{tabular}{lcc@{\hspace{5mm}}cc@{\hspace{5mm}}r@{\hspace{5mm}}r}
\hline
     ID & R.A.      & DEC      & $F606W$   & $F606W\!-\!F814W$ & pm$_{RA}$cos(Dec)         & pm$_{Dec}$ \\
         & (J2000)  & (J2000) & [mag] & [mag]    & [mas/yr] & [mas/yr] \\
\hline 
25341 & 48.0822474 & $-$55.2259794 & 18.285 & 0.53   & $-$0.0040 & 0.182   \\
29884 & 48.0734013 & $-$55.2210313 & 18.485 & 0.165  & $-$0.043  & $-$0.082  \\
30687 & 48.0726997 & $-$55.2208797 & 18.425 & 0.466  & 0.011   & 0.0010  \\
25466 & 48.0823863 & $-$55.2190322 & 17.57  & 0.139  & 0.091   & 0.06    \\
31914 & 48.0707694 & $-$55.2174152 & 18.425 & 0.083  & 0.012   & $-$0.039  \\
65969 & 48.0761793 & $-$55.2168233 & 18.468 & 0.147  & 0.013   & 0.0030  \\
70287 & 48.0693072 & $-$55.2160585 & 18.531 & 0.411  & $-$0.059  & 0.235   \\
76977 & 48.0592317 & $-$55.2158611 & 18.637 & 0.391  & 0.069   & 0.168   \\
67539 & 48.0739587 & $-$55.2157406 & 17.671 & 0.075  & 0.023   & $-$0.046  \\
81943 & 48.0498728 & $-$55.2149442 & 18.686 & 0.439  & $-$0.061  & $-$0.0090 \\
76627 & 48.0594802 & $-$55.214196  & 18.17  & 0.239  & 0.013   & 0.126   \\
72106 & 48.06711   & $-$55.2135349 & 17.846 & 0.107  & $-$0.041  & 0.055   \\
81984 & 48.0501694 & $-$55.2131204 & 18.525 & 0.172  & $-$0.019  & $-$0.141  \\
68014 & 48.0733391 & $-$55.2127941 & 18.323 & 0.032  & $-$0.0070 & $-$0.0090 \\
67728 & 48.0739974 & $-$55.2096186 & 18.182 & 0.526  & 0.13    & $-$0.191  \\
74845 & 48.0628007 & $-$55.2085887 & 18.243 & 0.2    & 0.011   & 0.0010  \\
73043 & 48.0655777 & $-$55.2073627 & 18.409 & 0.191  & $-$0.061  & 0.065   \\
8755   & 48.0655993 & $-$55.2368193 & 18.176 & 0.172  & 0.055   & 0.094   \\
32865 & 48.0687891 & $-$55.2225572 & 18.135 & 0.282  & 0.142   & $-$0.043  \\
34733 & 48.065501  & $-$55.2201113 & 18.044 & 0.18   & $-$0.061  & $-$0.02   \\
69581 & 48.0709143 & $-$55.2121116 & 18.053 & 0.366  & $-$0.02   & 0.107   \\
96886 & 48.0733873 & $-$55.2002784 & 17.787 & 0.19   & $-$0.0040 & 0.05    \\
5315   & 48.0808075 & $-$55.2349835 & 19.146 & 0.347  & 0.066   & 0.152   \\
12844 & 48.0469911 & $-$55.2331005 & 19.498 & 0.264  & $-$0.09   & 0.0070  \\
34871 & 48.0652367 & $-$55.2287238 & 18.898 & 0.249  & $-$0.014  & 0.075   \\
30492 & 48.0723216 & $-$55.2275608 & 19.315 & 0.338  & 0.157   & $-$0.0020 \\
32394 & 48.0693159 & $-$55.226576  & 18.926 & 0.105  & 0.01    & 0.139   \\
27018 & 48.0788222 & $-$55.2257367 & 19.796 & 0.346  & 0.013   & $-$0.088  \\
28759 & 48.0758134 & $-$55.2254191 & 18.87  & 0.232  & 0.03    & 0.073   \\
36170 & 48.0632261 & $-$55.224378  & 18.957 & 0.325  & 0.042   & 0.282  \\
\hline
\end{tabular}
\end{table*}

\section{Summary}
\label{txt:sum}
We present a comprehensive proper-motion analysis based on Hubble Space Telescope (HST) observations of \tr{38} Galactic GCs obtained by two public treasury surveys ("{\it The ACS Globular Cluster Survey}", GO-10775 and "{\it The HST Legacy Survey of Galactic Globular Clusters: Shedding UV Light on Their Populations and Formation}", GO-13297). We used WFC3/UVIS imaging data and apply the reduction routines developed by the STScI in order to derive precise astrometric catalogs of the inner $\sim\!3\arcmin\!\times\!3\arcmin$ GC regions. We combine those catalogs with calibrated ACS photometric catalogs and use these data to measure precise relative proper motions of stars in our sample GCs and perform a subsequent cluster membership selection. We study the accuracy of our proper motion measurements using estimates of the central velocity dispersion in each of our clusters and find very good agreement with previous similar studies in the literature. Finally, we construct homogeneously defined BSS selection criteria in order to derive proper-motion cleaned BSS catalogs in all \tr{38} GCs. The proper motion decontamination allows the unambiguous selection of \tr{yellow stragglers} in the CMD, which had been previously not possible in most of the past BSS studies given the high field star contamination in those CMD regions. These BSS catalogs and their proper motion information are being used by our group to study the dynamical state of BSSs in general and put them in the context of the GC dynamical evolution. The results from these studies will be soon presented in a forthcoming paper.

\section*{Acknowledgements}
We gratefully acknowledge support from CONICYT through the \textit{Doctorado Nacional} Fellowship, ALMA-CONICYT Project No.~37070887, FONDECYT Regular Grant No.~1121005, FONDAP Center for Astrophysics (15010003), and BASAL Center for Astrophysics and Associated Technologies (PFB-06), as well as support from SOCHIAS, the HeidelbergCenter in Santiago de Chile and the {\it Deutscher Akademischer Austauschdienst (DAAD)}. \tr{We thank the referee, Nathan Leigh, for very constructive comments and suggestions.} MS would also like to thank Andrea Bellini and Laura Watkins for their kind help in the understanding of proper motion analysis. All of the data presented in this paper were obtained from the Mikulski Archive for Space Telescopes (MAST). STScI is operated by the Association of Universities for Research in Astronomy, Inc., under NASA contract NAS5-26555. Support for MAST for non-HST data is provided by the NASA Office of Space Science via grant NNX09AF08G and by other grants and contracts. This research has made use the Aladin plot tool and the TOPCAT table manipulation software, found in: http://www.starlink.ac.uk/topcat/.









\bsp	
\label{lastpage}
\end{document}